\documentclass{aa}  

\usepackage{graphicx}
\usepackage{txfonts}
\usepackage{multirow}
\usepackage{colortbl}

\usepackage{hyperref}  
\hypersetup{colorlinks=true,linkcolor=[rgb]{1.,0.2,0.2},citecolor=[rgb]{0.1,0.4,1.},filecolor=[rgb]{0.7,0.2,0.2},urlcolor=[rgb]{0.7,0.2,0.2}}

\usepackage{color}
\definecolor{blue}{rgb}{0., 0., 1}

\newcommand{\hii}{\textrm{H}\textsc{ii}}
\newcommand{\hi}{\textrm{H}\textsc{i}}

\newcommand{\oiii}{[\textrm{O}\textsc{iii}]}

\newcommand{\oiidoub}{[\textrm{O}\textsc{ii}]\ensuremath{\lambda3727,3729}}

\newcommand{\oiiiv}{[\textrm{O}\textsc{iii}]\ensuremath{\lambda5007}}
\newcommand{\oiiiiv}{[\textrm{O}\textsc{iii}]\ensuremath{\lambda4959}}
\newcommand{\oiiialone}{[\textrm{O}\textsc{iii}]}

\newcommand{\oiiidoub}{[\textrm{O}~\textsc{iii}]\ensuremath{\lambda\lambda4959,5007}}
 
\newcommand{\ha}{\ifmmode {\rm H}\alpha \else H$\alpha$\fi}
\newcommand{\hb}{\ifmmode {\rm H}\beta \else H$\beta$\fi}
\newcommand{\lya}{\ifmmode {\rm Ly}\alpha \else Ly$\alpha$\fi}
\newcommand{\pg}{\ifmmode {\rm P}\gamma \else Pa$\gamma$\fi}
\newcommand{\lyb}{\ifmmode {\rm Ly}\beta \else Ly$\beta$\fi}
\newcommand{\lyg}{\ifmmode {\rm Ly}\gamma \else Ly$\gamma$\fi}

\newcommand{\nv}{\textrm{N}\textsc{v}\ensuremath{\lambda1240}}

\newcommand{\civmed}{\textrm{C}\textsc{iv}\ensuremath{\lambda 1550}}
\newcommand{\heii}{\textrm{He}\textsc{ii}\ensuremath{\lambda1640}}

\newcommand{\flyc}{\ifmmode  \mathrm{f}_\mathrm{esc}\mathrm{(LyC)} \else $\mathrm{f}_\mathrm{esc}\mathrm{(LyC)}$\fi}

\def\sfr{M$_{\odot}$~yr$^{-1}$}
\def\kms{km s$^{-1}$}

\def\ergs{\ifmmode \mathrm{erg\hspace{1mm}s}^{-1} \else erg s$^{-1}$\fi}

\def\micron{\ifmmode \mu\mathrm{m} \else $\mu$m\fi}
\def\msun{\ifmmode \mathrm{M}_{\odot} \else M$_{\odot}$\fi}
\def\msunyr{\ifmmode \mathrm{M}_{\odot} \hspace{1mm}{\rm yr}^{-1} \else $\mathrm{M}_{\odot}$ yr$^{-1}$\fi}
\def\zsun{\ifmmode Z_{\odot} \else Z$_{\odot}$\fi}
\def\lsun{\ifmmode L_{\odot} \else L$_{\odot}$\fi}
\def\mstar{\ifmmode \mathrm{M}_{\star} \else M$_{\star}$\fi}

\begin{document}

\titlerunning{Star clusters as reionizers}
\title{
High star cluster formation efficiency in the strongly lensed {\rm Sunburst} Lyman-continuum galaxy at z=2.37\thanks{Based on observations collected at the European Southern Observatory for Astronomical research in the Southern Hemisphere under ESO programmes ID 0103.A-0688(A), 0103.A-0688(C) (PI E. Vanzella). }}

\authorrunning{E. Vanzella et al.}
\author{
E. Vanzella \inst{\ref{inafbo}} \fnmsep\thanks{E-mail: \href{mailto:eros.vanzella@inaf.it}{eros.vanzella@inaf.it}},
M. Castellano \inst{\ref{inafrome}},
P. Bergamini \inst{\ref{inafbo},\ref{unimi}},
M. Meneghetti \inst{\ref{inafbo}},
A. Zanella \inst{\ref{inafpd}},
F. Calura \inst{\ref{inafbo}},
G. B. Caminha \inst{\ref{maxplanck}},\\
P. Rosati \inst{\ref{unife},\ref{inafbo}},
G. Cupani \inst{\ref{inafts}},
U. Me\v{s}tri\'{c} \inst{\ref{inafbo}},
G. Brammer \inst{\ref{NBI}},
P. Tozzi \inst{\ref{arcetri}},
A. Mercurio \inst{\ref{inafna}},
C. Grillo \inst{\ref{unimi}},
E. Sani \inst{\ref{ESO}},
S. Cristiani \inst{\ref{inafts}},\\
M. Nonino \inst{\ref{inafts}},
E. Merlin \inst{\ref{inafrome}},
and G.V.~Pignataro \inst{\ref{difabo}}
}

\institute{
INAF -- OAS, Osservatorio di Astrofisica e Scienza dello Spazio di Bologna, via Gobetti 93/3, I-40129 Bologna, Italy \label{inafbo} 
\and
INAF -- Osservatorio Astronomico di Roma, Via Frascati 33, I-00078 Monte Porzio Catone (RM), Italy \label{inafrome}
\and
Dipartimento di Fisica, Universit\`a  degli Studi di Milano, via Celoria 16, I-20133 Milano, Italy \label{unimi}
\and
INAF -- Osservatorio Astronomico di Padova, Vicolo Osservatorio 5, 35122, Padova, Italy \label{inafpd}
\and
Max-Planck-Institut f\"ur Astrophysik, Karl-Schwarzschild-Str. 1, D-85748 Garching, Germany \label{maxplanck}
\and
Dipartimento di Fisica e Scienze della Terra, Universit\`a degli Studi di Ferrara, via Saragat 1, I-44122 Ferrara, Italy \label{unife}
\and
INAF -- Osservatorio Astronomico di Trieste, via G. B. Tiepolo 11, I-34143, Trieste, Italy \label{inafts}
\and
Cosmic Dawn Center, Niels Bohr Institute, University of Copenhagen, Juliane Maries Vej 30, DK-2100 Copenhagen \O, Denmark\label{NBI} 
\and
INAF -- Osservatorio Astrofisico di Arcetri, Largo E. Fermi, I-50125, Firenze, Italy \label{arcetri}
\and
INAF -- Osservatorio Astronomico di Capodimonte, Via Moiariello 16, I-80131 Napoli, Italy \label{inafna}
\and
European Southern Observatory, Alonso de Cordova 3107, Casilla 19, Santiago 19001, Chile \label{ESO}
\and
DIFA -- Dipartimento di Fisica e Astronomia, Università di Bologna, via Gobetti 93/2, I-40129 Bologna, Italy \label{difabo} 
}

\date{} 

 
\abstract
{We investigate the strongly lensed ($\mu \simeq \times 10-100$) Lyman continuum (LyC) galaxy, dubbed {\rm Sunburst}, at $z=2.37$, taking advantage of a new accurate model of the lens. A  characterization of the intrinsic (delensed) properties of the system yields a size of $\simeq 3$ sq.kpc, a luminosity of M$_{\rm UV} = -20.3$, and a stellar mass of  M~$\simeq 10^{9}$ \msun; 16\% of the ultraviolet light is located in a 3 Myr old gravitationally bound young massive star cluster (YMC), with an effective radius of $\sim 8$ pc (corresponding to 1 milliarcsec without lensing) and a dynamical mass of $\sim 10^{7} $ \msun\ (similar to the stellar mass) -- from which LyC radiation is detected ($\lambda < 912$\AA). The star formation rate and stellar mass surface densities for the YMC are
Log$_{10}(\Sigma_{\rm SFR}[{\rm M}_{\odot}{\rm yr}^{-1}{\rm kpc}^{-2}])~\simeq 3.7$ 
and 
Log$_{10}(\Sigma_{\rm M}[{\rm M}_{\odot}{\rm pc}^{-2}])~\simeq 4.1$, with sSFR $> 330$ Gyr$^{-1}$, consistent with the values observed in local young massive star clusters. The inferred outflowing gas velocity ($>300~$\kms) exceeds the escape velocity of the cluster.
The resulting relative escape fraction of the ionizing radiation emerging from the entire galaxy is higher than 6-12\%, whilst it is $\gtrsim 46-93$\% if inferred from the YMC multiple line of sights. At least 12 additional unresolved star-forming knots with radii spanning the interval $3-20$ pc (the majority of them likely gravitationally bound star clusters) are identified in the galaxy.
A significant fraction (40-60\%) of the ultraviolet light of the entire galaxy is located in such bound star clusters. In adopting a formation timescale of the star clusters of 20 Myr,   a cluster formation efficiency $\Gamma \gtrsim 30$\%.
The star formation rate surface density of the {\rm Sunburst} galaxy (Log$_{10}(\Sigma_{\rm SFR}) = 0.5_{-0.2}^{+0.3}$) is consistent with the high inferred  $\Gamma$, as observed in local galaxies experiencing extreme gas physical conditions.
Overall, the presence of a bursty event (i.e., the 3 Myr old YMC with large sSFR) significantly influences the morphology (nucleation), photometry (photometric jumps), and spectroscopic output (nebular emission) of the entire galaxy. Without lensing magnification, the YMC would be associated to an unresolved 0.5~kpc$-$size star-forming clump. The delensed LyC and UV magnitude $m_{1600}$ (at 1600\AA) of the YMC are $\simeq 30.6$ and $\simeq 26.9$, whilst the entire galaxy has m$_{1600} \simeq 24.8$.
The {\rm Sunburst} galaxy shows a relatively large rest-frame equivalent width of EW$_{\rm rest}$(\hb + \oiiidoub)~$\simeq 450$\AA,\ with the YMC contributing to $\sim 30$\% (having a local EW$_{\rm rest} \simeq 1100$\AA) and $\sim 1$\% of the total stellar mass.
If O-type (ionizing) stars are mainly forged in star clusters, then such engines were the key ionizing agents during reionization and the increasing occurrence of high equivalent width lines (\hb+\oiiialone) observed at $z>6.5$ might be an indirect signature of a high frequency of forming massive star clusters (or high $\Gamma$) at reionization. Future facilities, which will perform at few tens milliarcsec resolution (e.g., VLT/MAVIS or ELT), will probe bound clusters on moderately magnified ($\mu < 5-10$) galaxies across cosmic epochs up to reionization.
}

   \maketitle
%

\section{Introduction}

A common morphological property of high redshift star-forming galaxies is the presence of clumps \citep[][]{elmegreen07,forser11, wuyts13,bournaud16,guo18,zanella15,zanella19,vanz_mdlf}. 
Strong gravitational lensing easily reveals such clumps down to a $\sim 100$ pc scale \citep[][]{livermore15, rigby17, cava18, mirka19, mirka17} and the sizes of massive stellar clusters ($ \lesssim 30$ pc) in high magnification regimes, $\mu > 20$ \citep{vanz19, vanz_sunburst, johnson17}. Probing the presence of clumps at the smallest physical scales remains crucial to understand the evolution of high-redshift galaxies and it is only recently that observations have been capable of achieving the resolution needed to detect the predicted substructures and thus understand their formation \citep[][and the references therein]{faure21, elmegreen20}.

Searching for the presence of single star clusters within high redshift clumps represents the next frontier and, certainly, a key science goal of future extreme adaptive optics facilities (XAO, eXtreme Adaptive Optics, e.g., \citealt[][]{vanz_mdlf}). 
Nowadays, gravitational lensing gives us the opportunity to access such small scales ($<100-200$ pc). The truth is, however, that even after relatively high amplification, $\mu<10-20$, it is still a challenge to probe scales smaller than a few tens of pc that are key for the identification of single star clusters. Therefore, higher magnification regimes are required, which, in turn, are more rare and difficult to handle as they are generally subject to high modeling uncertainties. Now, new-generation lensing models, based on an unprecedented number of constraints, are providing a significant advances in this direction \citep[e.g.,][]{bergamini21}, especially in the vicinity of (or on the) critical lines \citep[][]{vanz_popiii}. 

The system described in this work (dubbed {\rm Sunburst} in the following) belongs to the category of the super-lensed arcs, subject to amplification values spanning the range of $15 <\mu< 100$.
{\rm Sunburst} is a strongly magnified galaxy at redshift 2.37 discovered by \citet{dahle16}, gravitationally lensed by the Planck cluster PSZ1 G311.65-18.48 (at $z=0.44$) and split over four very bright arcs (with the brightest arc having magnitude R~$\sim18$). 
Several multiple images (more than 50 in total) of the same galaxy and internal star-forming knots populate the giant arcs.
Among them, a compact star-forming region, identified 12 times and showing a structured \lya\ profile with multi-peaks and with non-zero emission at the systemic redshift, was identified as a candidate region with Lyman continuum leakage (optically thin to ionizing photons, LyC, $\lambda <912$\AA, \citealt{rivera17}). Subsequent observations with Hubble confirmed the LyC emerging from the same 12 multiple regions, in agreement with the exceptionally high image multiplicity \citep[][]{rivera19}. \citet{chisholm19} accurately estimated an age of 3 Myr, sub-solar metallicity of 0.5 Z$_{\odot}$ and E(B-V)=0.15, while \citet{vanz_sunburst} derived the dynamical age from the limits on the stellar mass and size of the system (as a function of magnification), recognizing it as a plausible massive young star cluster with a stellar mass of $\sim 10^{7}$~\msun, an effective radius smaller than 25 pc, and showing evident spectral signatures of massive O-type and Wolf-Rayet stars from the VLT/MUSE spectrum. 
The resulting initial constraints on the star formation rate and stellar mass surface densities of the LyC emitting knot also resemble those of local gravitationally bound star clusters \citep[][see their Sect.~4.1.1]{vanz19}. 
Sources like {\rm Sunburst} are unique laboratories for improving our understanding of the physical mechanisms related to escaping radiation at cosmological distances. This makes them very useful in the  search for ionizing sources at earlier epochs during cosmic reionization ($z>6$), when the LyC radiation cannot be observed directly due to high cosmic opacity rapidly increasing in the first $\simeq 1.5$ Gyr \citep[][]{worseck14, romano2019}. They are also key for making comparisons with LyC emitters identified in the Local Universe (including candidates), considered analogs of those with high redshift \citep[][]{izotov21loz_analogs,schaerer16,jaskot19,Wang_[SII]_21}.

An accurate model of the lens is required to unveil the intrinsic properties of such sources; this type of model is described in a companion paper \citep[][]{pignataro21}. 
Here, we present refined estimates of the physical quantities of the LyC star cluster, including the other knots identified in the surrounding, and we derive the global properties of the host galaxy. 

This work is organized as follows. Section~\ref{data} describes the data used and Sect.~\ref{details} presents the {\rm Sunburst} lensed system, the lens model, the emergence of stellar clusters, and an estimate of the dynamical mass for the stellar cluster showing Lyman continuum emission. Section~\ref{sec:sunburst} describes the galaxy hosting the aforementioned star clusters, along with its intrinsic appearance. In Sect.~\ref{sec:CFE}, we compute the cluster formation efficiency and in Sect.~\ref{sec:fesc}, we compute  its global (along line of sight) escape fraction of ionizing photons. Section~\ref{sec:summary} summarizes our results and describes the implications drawn from possible analogies with  sources at $z>6.5$ that are likely responsible for reionization (Sect.~\ref{CFE_EOR}). We also present the current limitations and future prospects in that section (Sect.~\ref{future}).

We assume a flat cosmology with $\Omega_{M}$= 0.3,
$\Omega_{\Lambda}$= 0.7 and $H_{0} = 70$ km s$^{-1}$ Mpc$^{-1}$. With this assumption, one arcsec at $z=2.37$ corresponds to a projected physical scale of 8200 parsec. Unless specified, all magnitudes are given in the AB system.

\section{Data}
\label{data}

The galaxy cluster PSZ1 G311.65-18.48 was observed in several bands with the Hubble Space Telescope between 2018-2020 under the programs 15101 (PI Dahle), 15949 (PI Gladders), and 15377 (PI Bayliss). We retrieved the WFC3/F390W (total integration time 5.8 ks), WFC3/F410M (13 ks), WFC3/F555W (5.6 ks), WFC3/F606W (5.9 ks), ACS/F814W (5.3 ks), WFC3/F098M (1.4 ks), WFC3/F105W (1.3 ks), WFC3/F140W (2.7 ks), and WFC3/F160W (1.3 ks) exposures from the MAST archive. We processed them with the \textsc{grizli} software\footnote{https://www.github.com/gbrammer/grizli} following the procedure described by \cite{williams21}.  Each of the groups of exposures in a given filter and epoch (a ``visit'' in the \textit{Hubble} observation nomenclature) was aligned to the GAIA DR2 absolute astrometric frame \citep{gaia2016, gaia2018} and accounting for proper motions.  The PSZ1 G311.65-18.48 field has a high density of GAIA sources and the absolute astrometric precision of all image mosaics is $\sim$10~milliarcsec (1$\sigma$).  The morphological analyses below are performed on mosaics in each filter drizzled to a common 30~milliarcsec pixel grid with the \textsc{DrizzlePac} software \citep{gonzaga12}.

Additional ground-based observations obtained with the VLT/X-Shooter (PI Vanzella) are used in this work and presented in Appendices~\ref{ha} and~\ref{dyn_mass}. A comprehensive analysis of the full spectrum of {\rm Sunburst}, from \lya\ to \ha\ wavelengths, will be presented in a forthcoming work.
VLT/X-Shooter data acquisition and reduction on {\rm Sunburst} have already been presented in \citet{vanz20_laser} and described in Appendix~\ref{ha}, where we focus on key observables useful for the analysis described in this work.

MUSE observations in the narrow field mode configuration of a portion of the lensed system are also presented in this work (PI Vanzella, see Appendix~\ref{appendix_NFM}),
by highlighting the exceptional performances of the extreme adaptive optics, as well as some limitations in cases such as the lensed systems described here.

\section{Anatomy of the {\rm Sunburst}: Detecting star clusters at $z=2.37$}
\label{details}

\subsection{Lensing model: A very amplified system}
\label{thesystem}

The work by \citet{pignataro21} presents the details of the lens model, along with the
predicted positions of all 62 multiple images of various galaxies and the associated families currently confirmed in the redshift range of $1<z<3.5$ (out of which 54 belongs to the {\rm Sunburst} galaxy at z=2.37).
The high level of accuracy achieved for this new model is underscored by its ability to predict the observed positions of 62 multiple images with a rms error of only $0.14''$.

 \begin{figure*}
        \centering
        \includegraphics[width=\linewidth]{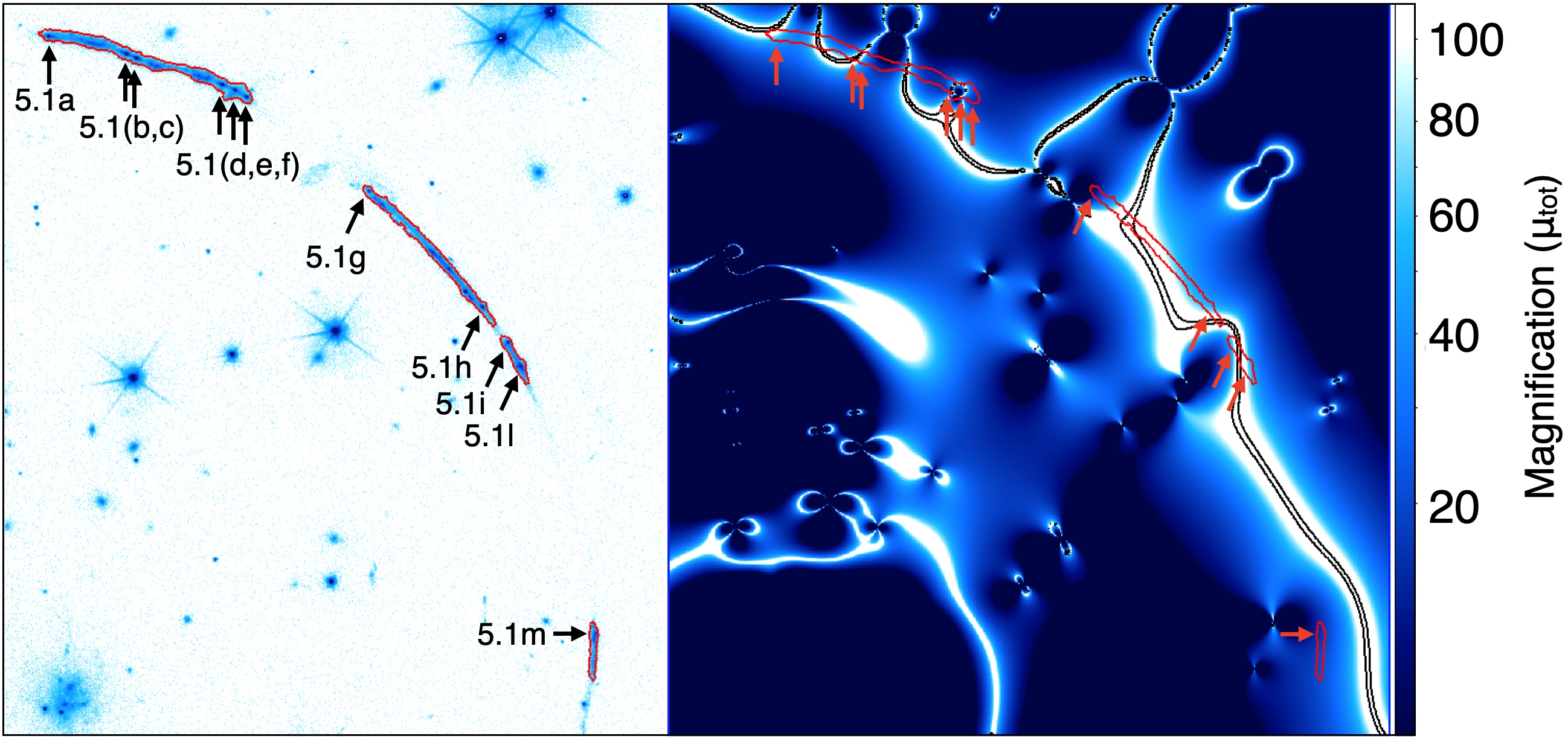}
        \caption{Panoramic view of the main arcs discussed in this work. Left panel (in F555W-band image): Arcs are marked with red contours to guide the eye. The arrows indicate 11 multiple images of the star-forming knots labeled 5.1$a-m$ (see text and Figure~\ref{pano} for additional details). The magnification map (total magnification, $\mu_{tot}$) is reported on the right panel (from \citealt{pignataro21}) coded according with the color bar in square root scale: it saturates at $\mu_{tot} \simeq 10$ (black) and 100 (white). The black contours enclose the region where $\mu_{tot}>500$, marking the locus of the critical line. The arcs contain multiple images of the same portions of the {\rm Sunburst} galaxy and are subjected to amplification spanning the range $15-100$, with the critical line crossing the arcs several times. 
        }
        \label{lens_map}
\end{figure*}

\begin{figure*}
        \centering
        \includegraphics[width=\linewidth]{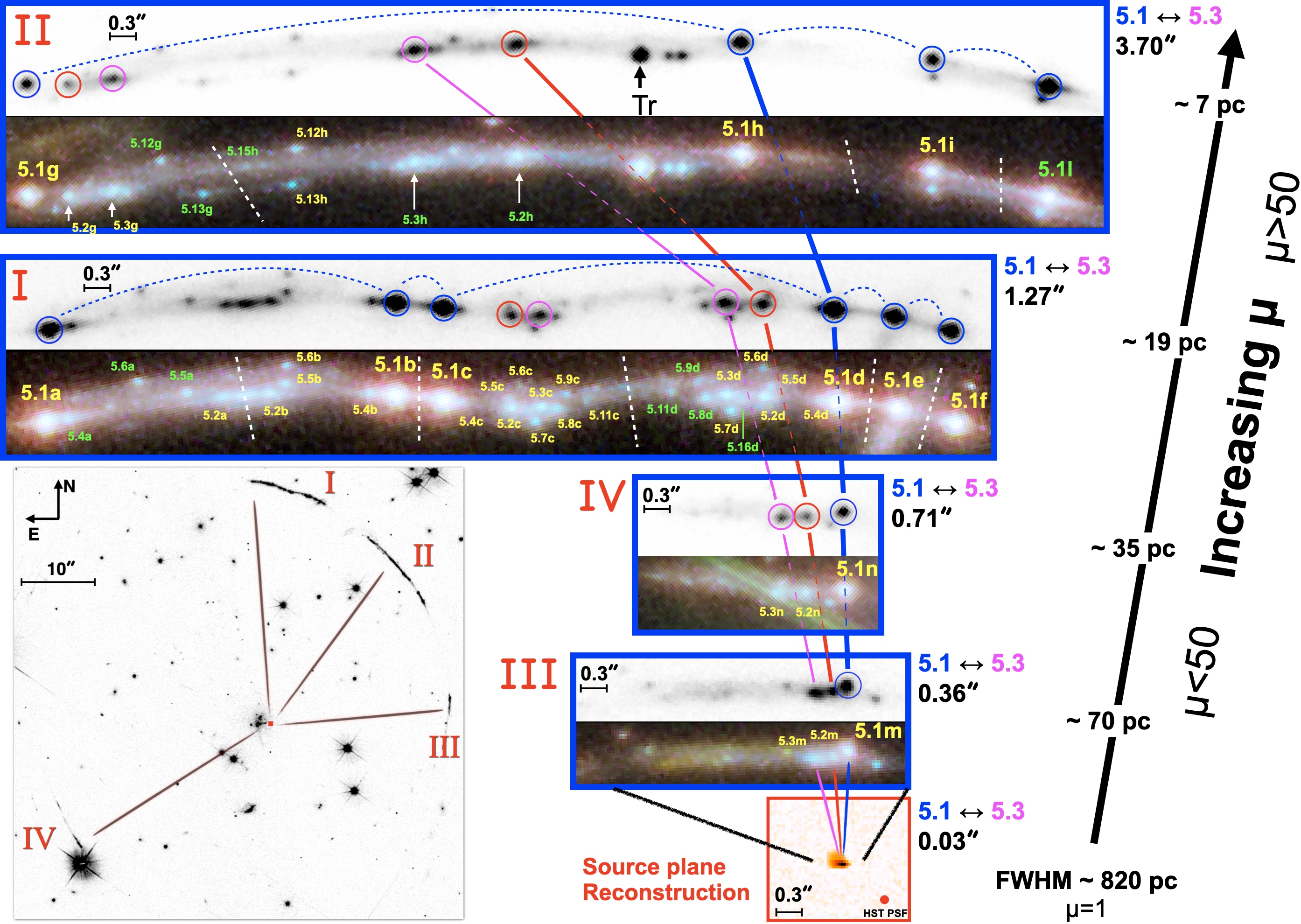}
        \caption{
        Identification of the multiple images of compact star-forming knots is shown. In the bottom-left of the figure, the negative gray-scaled HST ACS/F555W image shows the panoramic view of the four arcs, labeled as I, II, III, and IV, with the little red square in the middle indicating the source region subject to large amplification. The SF knots are shown in the blue border insets labeled as I, II, III, and IV, following the arcs notation in the bottom left inset and with increasing magnification from bottom to top (long arrow on the right). Each inset includes the negative gray scale F555W image and a color image obtained by combining the WFC3/UVIS F275W (blue), WFC3/UVIS F606W (cyan), ACS/WFC F814W (yellow), and WFC3/IR F160W (red) filters (ESA/Hubble, NASA, Rivera-Thorsen et al., CC BY 4.0). 
        Star-forming regions are indicated following the nomenclature of \citep{pignataro21}. White dashed lines mark the position of the critical lines. The main triplet, 5.1, 5.2, and 5.3 are indicated with blue, red, and magenta open circles, respectively. The objects labeled with green IDs are the knots reported in Table~\ref{knots}. Those marked in yellow show all the identified multiple images of the knots. The inset in the bottom right (red contour) shows the reconstructed image of the {\rm Sunburst} galaxy on the source plane, where the triplet 5.1, 5.2, and 5.3 would merge into a single HST pixel ($\sim 30$ mas separation). The observed angular separation between the knots 5.1 and 5.3 is reported on the right side of each inset, and the typical physical scales subtended by the FWHM ($0.1''$) along tangential direction is shown over the long black arrow to the right (arc II includes star-forming region smaller 10 pc).
        The transient (``Tr'') discussed in \citet{vanz20_laser} and corresponding to the knot 5.10 is indicated with a black arrow in the inset II.
        }
        \label{pano}
\end{figure*}

We took advantage of this lens model to recognize each mirrored and multiply imaged star-forming knot of {\rm Sunburst}.
The geometry of the lens shows a rather complicated behavior of the critical lines at the redshift of {\rm  Sunburst} due 
to several perturbers placed along the line of sight, which 
add to the underlying mass distribution of the galaxy cluster. Such a configuration boosts the number of multiple images and total magnification values ($\mu_{tot}$), with $\mu_{tot}$ spanning the range between 15 and 100. 
Figure~\ref{lens_map} shows the magnification map along with the position of the relevant arcs discussed in this work.
For example, one of the knots (labeled 5.1 in Figure~\ref{lens_map} and  Figure~\ref{pano}) is detected 12 times due to a combination of the galaxy cluster and perturbers. The set of multiple images of a single knot is called a "family" and each member of the family is denoted with a different letter. For example, each of the 12 images of 5.1, 5.1(a$-$n) is subjected to a different magnification.  
Along the giant arcs, the total magnification (denoted as $\mu_{tot}$) is dominated by tangential stretch, such that the ratio between the total magnification $\mu_{tot}$ and the tangential one, $\mu_{tang}$, is nearly constant and close to $1.3 - 1.4$. The ratio $\mu_{tot}/ \mu_{tang}$ is referred to the radial magnification, which is therefore very low on the arcs.
Figure~\ref{pano} shows the panoramic view of the four arcs: I, II, III, and IV that contain several portions of the same galaxy. 
Thirteen individual SF knots 
replicate several times with different amplification values, according to the complex geometry of the lens. 
The list of knots is reported in Table~\ref{knots},\ presenting those highlighted in green labels in Figure~\ref{pano}.

Statistical errors on magnification have been calculated by extracting the total(tangential) magnification $\mu_{tot}$ ($\mu_{tang}$) at the model predicted positions (of the star-forming knots discussed below) over 500 realizations of the lens model by sampling the posterior probability distribution function. This has been performed by using a Bayesian Markov chain Monte Carlo (MCMC) technique \citep[as described in][]{bergamini21}. Total ($\mu_{tot}$) and tangential ($\mu_{tang}$) magnifications and the associated errors are reported in Table~\ref{knots}. The physical quantities reported in Table~\ref{knots} are calculated based on the best-fit lens model and the uncertainties over the grid of 500 magnifications. Overall, such large magnifications imply that a scale of a few tens of pc  can be probed along the tangential stretch,
allowing us to identify single star clusters (discussed in the next section).

\subsection{Emergence of star-forming regions at parsec scale}
\label{anatomy}

To provide the most stringent constraint on the size of each knot, we consider the brightest one within the associated family, characterized by a high magnification.
All the identified multiple images
of the knots, except for knot 5.1, appear as unresolved objects in the F555W-band (1600\AA\ rest-frame), independently of their magnification value. The F555W is the bluest band probing the ultraviolet continuum (not including \lya\ emission or IGM attenuation). Under these conditions, this band is also the one with the narrower PSF, thus, it offers the best probe (spatial contrast) of the knots and therefore enables optimal estimates of the sizes in the ultraviolet. 
The current HST data in the near infrared bands do not allow us to address the sizes in the optical rest-frame with the same level of detail. As discussed in Sect.~\ref{future}, this will be  performed with ease thanks to future facilities based on XAO.

\begin{figure}
        \centering
        \includegraphics[width=\linewidth]{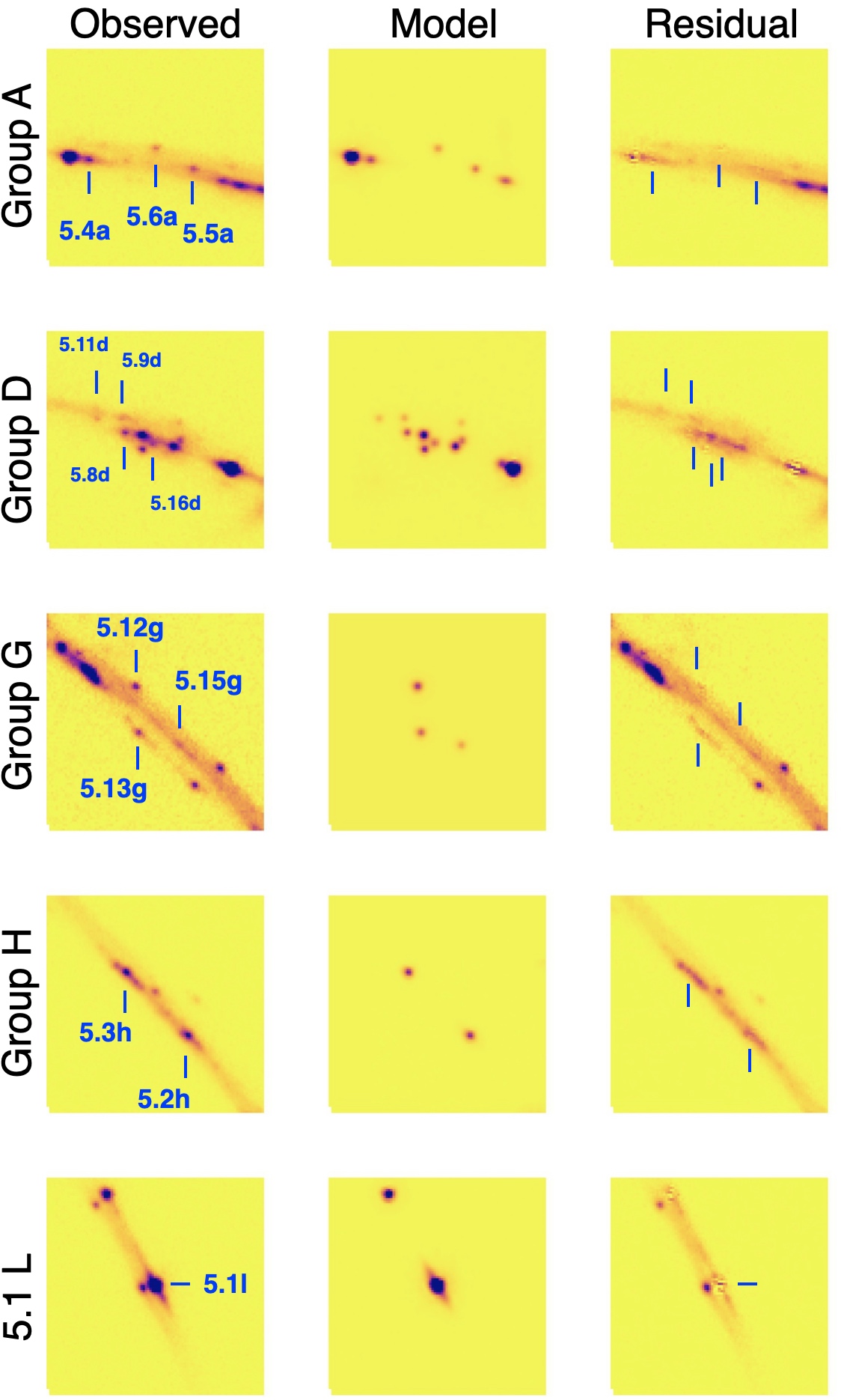}
        \caption{{\tt Galfit} fitting of the star-forming knots reported in Table~\ref{knots} zooming on various regions (groups A, D, G, H and L, see Figure~\ref{pano}). The left column shows the original F555W image, in which only the knots listed in Table~\ref{knots} are marked with their IDs. The middle column shows the {\tt Galfit} models and the right column shows the residual after subtracting the models, with the positions of the knots indicated as in the left column. Besides the IDs in the reference sample (marked in blue), additional multiply imaged knots have been modeled, such as the cloud of clusters of the group D (see the same group described in Figure~\ref{pano}).
        All knots are not resolved with resulting effective radii $\lesssim 1$ pixel, independently from the S\'ersic index ($n=0.5-4$).}
        \label{galfit}
\end{figure}

We modeled the two-dimensional light distribution with {\tt Galfit}.
{\tt Galfit} fitting has been performed by assuming the S\'ersic index $n=0.5$ (Gaussian), 1.0 (exponential) up to $n=4$ (de Vaucouleurs) light profiles.
Given the circular symmetric shapes of the knots, we fix the axis ratio ($b/a$) and position angle to 1 and 0, respectively. The inferred sizes always converge to compact solutions, independently of the adopted S\'ersic index, formally with PSF-deconvolved radii R$_{\tt eff} < 1$ pixel for all of them (1 pix = $0.03''$), but one, the brightest 5.1l which has an R$_{\tt eff}=2.0 \pm 0.6$ pix. The modeled images and residuals of the modeling (observed~$-$~model) are shown in Figure~\ref{galfit}.
Table~\ref{knots} summarizes the inferred intrinsic magnitudes resulting from the fit and the physical sizes (effective radii), along with the magnifications and associated uncertainties (statistical errors). 
The estimated radii span the range $<22$ pc, which likely represents upper limits (given the sources are unresolved), approaching sizes of a few pc ($<10$ pc) for the cases very close to the critical lines.
It is worth noting that all knots show a point-like shape also in the F390W band, implying a half width at half maximum (HWHM) of $0.04''$ (as inferred from stars in the field), which is consistent with the small upper limits on the effective radii derived above with {\tt Galfit} in F555W-band.

\subsection{Cocoon of gravitationally bound star clusters}
\label{clusters}
The small sizes inferred from modeling (R$_{\tt eff}\lesssim 22$ pc) might suggest that 
such star-forming knots are stellar clusters. In order to investigate such a possibility, we consider whether such compact star-forming knots
are gravitationally bound systems by calculating the dynamical age, $\Pi$
(equal to the ratio of the age and crossing time). 

The dynamical age $\Pi$ can be calculated as described in \citet{gieles11} (see also \citealt{adamo20} and \citealt{vanz_mdlf}). In particular, 
the crossing time expressed in Myrs is defined as ${\rm T_{CR} = 10 \times (R_{\tt eff}^{3} / GM)^{0.5}}$, where M and ${\rm R_{eff}}$ are the stellar mass and the effective radius, respectively, with ${\rm G \approx 0.0045~pc^{-3} \msun^{-1} Myr^{-2}}$ is the gravitational constant. Stellar systems evolved for more than a crossing time have ${\rm \Pi > 1}$, suggestive of being bound (\citealt{gieles11}). This criterion has been used extensively for the identification of star clusters in the local Universe  \citep[e.g.,][]{calzetti15,adamo17,ryon17}. It is valid under the following assumptions: the system is in virial equilibrium, follows a Plummer density profile, and the light traces the underlying mass.
Therefore the calculation of $\Pi$ for each  knot requires the knowledge of three main parameters: age, stellar mass, and effective radius.

\noindent {\it Age}. The estimation of recent star-forming events on short timescales would require age indicators such as the Balmer lines, which are highly uncertain $-$ despite the high magnification $-$ since a detailed, spatially resolved spectroscopy of each star-forming knot is not yet achievable.
On the other hand, the SED-fitting weakly constrains young ages (being essentially dependent on the assumed star formation histories, e.g., \citealt{carnall19}) and is affected by blurring in the near infrared bands, where the PSF degrades and the emission from the knots and the host galaxy cannot be easily deblended. Nonetheless, an estimate of the average ages comes from VLT/X-Shooter observations \citep[][]{vanz20_laser}.
As mentioned above, individual knots cannot be probed due to the relatively coarse resolution. However, we estimated the \ha\ equivalent width of a group of 10 knots (group "d", Figure~\ref{pano}), obtaining spatially-averaged limits on the age, assuming that stars formed with an instantaneous burst (see Appendix~\ref{ha}). For knot 5.1, the \ha\ emission is prominent with respect to the underlying continuum, implying an age younger than 5 Myr and consistent with the 3 Myr inferred by \citet{chisholm19}. For the other knots of the group "d" instead, the \ha\ EW smaller than 100\AA\ rest-frame suggests an average age larger than $\simeq 7$ Myr. In the following, we adopt an age of 3 Myr for 5.1 \citep[][]{chisholm19} and assume the other knots are older than 7 Myr (see Figure~\ref{zoomknots} and Table~\ref{knots}). As discussed below, we note that even when relaxing the ages to smaller values, $\Pi$ still remains close to 1 or higher. 
Moreover, it is worth noting that assuming different SF histories (continuous or declining) would result in older ages of the knots \citep[][]{zanella15}, further increasing the dynamical age $\Pi$.

\noindent \noindent {\it Stellar mass}.
For the same reasons described above, the estimation of the stellar mass from SED fitting suffers from the resolution issue on PSF-matched images. In addition, the apertures   the photometry is calculated within, on the basis of multiple images of the same family, are (lens) model-dependent.
This makes the choice of the aperture shape over multiple images tricky. Such an issue will need dedicated simulations with forward modeling techniques, which will be described in a forthcoming work. In this work, we rely on the ultraviolet F555W-band image, which has a narrow PSF (of $0.1''$), such that the majority of knots are well recognized (Figure~\ref{pano}).

We assume all the knots formed through instantaneous bursts (as is the case for 5.1). We derive the stellar masses by comparing the dust-corrected UV luminosity with Starburst99 models \citep[][]{leitherer14}. We adopt a dust extinction of E(B-V)=0.15 as derived by \citet{chisholm19} for knot 5.1 (corresponding to A1600~$\simeq 1$, \citealt{reddy16}), whereas for the remaining knots we adopt the average E(B-V) inferred from SED fitting (see below) of all multiple images of the same family. In fact, while the colors extracted from the PSF-matched images are preserved, the derivation of the stellar masses from SED fitting
is strongly limited by the aforementioned issues \citep[see also][]{cava18}.
The stellar masses have therefore been inferred from the dust-corrected F555W luminosity, assuming the aforementioned ages (3 Myr for 5.1 and 7 Myr for the rest), and span the interval $10^{5-7}$\msun, weakly dependent on the assumed metallicity: they vary by $\simeq 0.18$ dex if the metallicity spans the interval of $Z=0.001-0.02$ (values reported in Table~\ref{knots} assume  Z=0.008, with Z=0.02 indicating solar metallicity in the Sunburst99 nomenclature).
It is worth noting that the UV-based stellar mass for 5.1 is consistent with the inferred dynamical mass discussed below (Sect~\ref{dynamo}).
Despite the aforementioned limitations related to crowding and aperture size definition, we check the consistency between the stellar mass inferred for knot 5.1 with both SED fitting and the single ultraviolet band F555W. 
In fact, the expected biases mentioned above (i.e., the crowding and lens-dependent aperture photometry among multiple images tends to be reduced) if the knot is bright and compact, therefore dominating the signal almost independently from the chosen aperture. 
The two methods have been applied to eight multiple images of the same knot 5.1 (a,b,c,h,i,l,m,n).
Knots 5.1d,e,f,g have been excluded from the computations because affected by a foreground perturber which might bias the estimate of the magnification and affects the photometry of the images \citep[][]{pignataro21}. SED fitting has been performed on $0.2''$ diameter apertures, adopting a metallicity of 0.4Z$_{sun}$ (the value in the BC03 library closer to the metallicity of the youngest star cluster 5.1, which has likely inherited the metallicity of the host galaxy), E(B-V) in the range $0-1.1$ and age $<50$ Myr. The extracted quantities (e.g., luminosity and stellar masses) are compatible within multiple images and comparable among the two methods within the relevant uncertainties (the median stellar mass inferred from single 1600\AA\ band and SED fitting are Mass$(\rm UV) = 1.1(\pm0.5)\times 10^{7}$~\msun\ and Mass$(\rm SED) = 0.9(\pm0.6)\times 10^{7}$~\msun, respectively.

\noindent \noindent {\it Sizes}. As described in Sect.~\ref{anatomy}, we verify the compactness of the knots by fitting the light profile with {\tt Galfit}. We obtain upper limits on the radii smaller than the half width at half maximum.
All of them have radii smaller than 22pc, whereas 5.1 is marginally resolved with a radius of $\simeq 8.5$ pc.

All the quantities with uncertainties are reported in Table~\ref{knots} along with the dynamical ages. The resulting dynamical age $\Pi$ exceeds 1 (gravitationally bound) for the majority of the knots, suggesting they indeed are likely to be bound stellar clusters. It is worth calculating for each of them the minimum age above which $\Pi$ exceeds 1. Since the UV-based stellar mass depends on the assumed age, we recalculate the mass at each adopted age, starting from 1 Myr and increasing it by 0.1 Myr till the condition $\Pi=1$ is reached. The last column of Table~\ref{knots} reports the age and stellar mass at $\Pi=1$. Even in the case of younger ages (1-3 Myr) most of the knots are still bound. Therefore, the age limit ($>7$ Myr) based on the \ha\ equivalent width discussed above indicates that all knots are likely bound star clusters, including the knot 5.1, which is the most extreme case of a 3-Myr-old YMC with detected LyC leakage.

%
\begin{table*}
\caption{Summary of the identified SF knots observed on {\rm Sunburst}.}
\label{knots}      
\centering          
\begin{tabular}{l c c c c c c c c | c }     
\hline\hline  
(\#1) & (\#2) & (\#3) & (\#4) & (\#5) & (\#6) & (\#7) & (\#8) & (\#9) & (\#10) \\
ID    &   M$_{\rm UV}~(\sigma)$ & E(B-V) &Age          & R$_{\rm eff}~(\sigma)$ & Mass         &  $\mu_{tot}(\sigma)$ & $\mu_{tang}(\sigma)$  & $\Pi~(\sigma)$ & Age - Mass\\ 
      &  \small{intr.}  &  & \small{Myr} & \small{pc} & \small{$\times 10^{6}$~M$_{\odot}$} &             &               &   \small{(at \#4)}        & \small{(at $\Pi=1$)}  \\
\hline
5.1a  &  -18.31(0.50)  &  0.15 & $3$  & 8.1(3.1)  &  9.1  & 74(59)  &  60(48)  &  2.6(4.7)  &  1.1 - 9.6 \\ 
5.1b  &  -18.43(0.32)  &  0.15 & $3$  & 8.2(3.2)  &  10.2  & 80(29)  &  60(22)  &  2.8(2.0)  &  1.1 - 16.2 \\ 
5.1c  &  -18.54(0.37)  &  0.15 & $3$  & 9.6(4.3)  &  11.3  & 66(29)  &  51(23)  &  2.3(1.8)  &  1.1 - 17.9 \\ 
5.1h  &  -18.63(0.12)  &  0.15 & $3$  & 10.8(3.1)  &  12.2  & 61(6)  &  45(4)  &  2.0(0.9)  &  1.4 - 18.6 \\ 
5.1i  &  -17.68(0.18)  &  0.15 & $3$  & 7.8(2.4)  &  5.1  & 84(13)  &  63(10)  &  2.1(1.8)  &  1.3 - 7.8 \\ 
5.1l*  &  -18.58(0.16)  &  0.15 & $3$  & 8.5(2.4)  &  11.7  & 76(12)  &  57(9)  &  2.8(2.6)  &  1.1 - 16.7 \\ 
5.1m  &  -18.68(0.08)  &  0.15 & $3$  & $<$19.5(9.6)  &  12.8  & 16(1)  &  13(1)  &  0.8(1.2)  &  3.5 - 16.6 \\ 
5.1n  &  -18.99(0.07)  &  0.15 & $3$  & $<$20.4(10.0)  &  17.1  & 15(1)  &  12(1)  &  0.9(1.3)  &  3.4 - 19.7 \\ 
\hline
5.2h*  &  -17.02(0.13)  &  0.06 & $\gtrsim 7$  & 4.8(0.6)  &  5.4  & 67(6)  &  51(5)  &  10.5(1.0)  &  1.2 - 2.8 \\ 
5.3h*  &  -18.09(0.23)  &  0.06 & $\gtrsim 7$  & 23.7(0.8)  &  14.7  & 16(21)  &  10(14)  &  1.6(2.1)  &  1.2 - 1.0 \\ 
5.4a*  &  -16.51(0.33)  &  0.05 & $\gtrsim 7$  & 5.4(1.2)  &  3.2  & 56(26)  &  46(21)  &  6.8(3.2)  &  1.3 - 1.1 \\ 
5.5a*  &  -16.04(0.15)  &  0.11 & $\gtrsim 7$  & 7.9(1.1)  &  3.0  & 38(5)  &  31(4)  &  3.7(0.5)  &  3.5 - 1.1 \\ 
5.6a*  &  -15.83(0.16)  &  0.15 & $\gtrsim 7$  & 9.0(1.4)  &  3.2  & 33(4)  &  27(4)  &  3.1(0.4)  &  4.0 - 1.5 \\ 
5.8d*  &  -17.05(0.13)  &  0.10 & $\gtrsim 7$  & 20.2(2.4)  &  7.1  & 18(2)  &  12(1)  &  1.4(0.1)  &  6.1 - 5.7 \\ 
5.9d*  &  -16.21(0.13)  &  0.15 & $\gtrsim 7$  & 21.4(2.5)  &  4.5  & 17(2)  &  11(1)  &  1.0(0.1)  &  7.1 - 4.6 \\ 
5.11d*  &  -15.54(0.13)  &  0.15 & $\gtrsim 7$  & 15.0(1.8)  &  2.4  & 24(3)  &  16(2)  &  1.3(0.1)  &  6.2 - 1.9 \\ 
5.16d*  &  -17.03(0.13)  &  0.06 & $\gtrsim 7$  & 22.0(2.8)  &  5.5  & 17(2)  &  11(1)  &  1.1(0.1)  &  7.2 - 6.4 \\ 
5.12g*  &  -14.58(0.17)  &  0.15 & $\gtrsim 7$  & 2.9(0.5)  &  1.0  & 110(14)  &  86(11)  &  9.7(1.3)  &  1.3 - 0.5 \\ 
5.13g*  &  -12.81(0.34)  &  0.04 & $\gtrsim 7$  & 0.9(0.4)  &  0.1  & 372(97)  &  282(74)  &  18.6(4.8)  &  1.1 - 0.1 \\ 
5.15h*  &  -11.73(0.24)  &  0.10 & $\gtrsim 7$  & 1.3(0.7)  &  0.1  & 484(25)  &  375(19)  &  7.2(0.4)  &  4.6 - 0.1 \\ 
\hline
$\dagger$HostM & -20.30(0.10) & 0.03 & 126 & $\sim 1000$ & 1000 & 19(3) & 12(3) & $-$  & $-$\\
\hline \hline
\end{tabular}
\tablefoot{Column \#1: ID of the knots reported in Figure~\ref{pano}; \#2: delensed magnitudes at 1600\AA; \#3: Adopted E(B-V): 0.15 for knot 5.1 as derived by \citet{chisholm19}, and the mean value inferred from the SED fitting for the other knots (see text for details); \#4: Assumed ages in Myr. The age of 3 Myr has been adopted for 5.1 \citep[][]{chisholm19}, and ages $\gtrsim 7$ Myr is assumed for the others. \#5: delensed effective radii at 1600\AA\ as derived from {\tt Galfit} modeling. Only 5.1 is resolved, whereas the radii of the other knots are upper limits; \#6: stellar mass in units of $\times 10^{6}$ \msun\ derived at the age reported in column \#3, accordingly to the adopted instantaneous burst of Starburst99 models, $Z=0.008$ Z$_{\odot}$ and Salpeter IMF with M$_{\rm UP}=100$~\msun. Being the ages lower limits, also the stellar masses are formally lower limits; \#7: total magnification from the best solution of the lens model (see Sect.~\ref{thesystem}); \#8: tangential magnification from the best solution of the lens model (see Sect.~\ref{thesystem}); \#9: dynamical age adopting the ages (\#4), stellar mass (\#6) and size (\#5) (see text for details). The limits on sizes and stellar masses imply such values tend to be lower limits; \#10: age and stellar mass at $\Pi = 1$, given with the same units as columns \#4 and \#6, respectively;
($\dagger$) Host galaxy parameters as derived from the SED-fitting procedure described in Sect.~\ref{sect:SEDandMORPH}; see also Figure~\ref{sed} where the 68\% central confidence interval is reported; (*) individual knots used in this manuscript to infer the cluster formation efficiency. Results from the multiple images of knot 5.1(1,b,c,h,i,l,m,n) are shown in the top part of the table. Knots 5.7c,d and 5.10 are multiple images of a possible stellar transient object describe by \citet{vanz20_laser} and are not included here. The knot 5.14 is also not included in this work since it is currently tentatively associated with multiple images of the knot 5.4.
}
\end{table*}

\subsection{LyC emitting cluster}

\subsubsection{Dynamical mass estimate}
\label{dynamo}

We estimated the dynamical mass of the knot labeled 5.1 by taking advantage of the observed \oiiiv\ line and the inferred velocity dispersion from high spectral resolution X-Shooter spectroscopy, R~$\simeq$~5600 (\citealt{rhoads14}, and see also Sect.~4.4 of \citealt{vanz_paving}).
The line profile and double Gaussian component fitting is shown in Figure~\ref{OIIIline} and the resulting dynamical mass of $10^{7}$ \msun\ is discussed in Appendix~\ref{dyn_mass}.  The resulting ratio between the stellar (photometry-based) and the dynamical masses is $\simeq 1$ and weakly depends on the magnification $\mu$. This is due to the fact that the photometric and dynamical masses scale with total and tangential magnifications, $\mu_{tot}$ and $\mu_{tang}$, respectively. In the case of {\rm Sunburst}, the tangential magnification largely dominates over the radial one, $\mu_{rad}$, such that $\mu_{tot} / \mu_{tang} = \mu_{rad} \simeq 1.3-1.4$, nearly constant along the arcs. The mass ratio, therefore, only includes $\mu_{rad}$, which is well constrained (with $< 10\%$ uncertainty, being far from the radial critical line).
The resulting stellar and virial masses are $1.1\times10^{7}$ \msun\ and $10^{7}$ \msun, respectively.
This indicates that the total mass is dominated by the stars, as was also found in most local bound clusters \citep[e.g.,][]{McLaughlin05}.

The combination of the stellar mass ($10^{7}$ \msun), age  (3 Myr) and size (R$_{\tt eff} \simeq 8$~pc) implies a star formation rate ($\Sigma_{\rm SFR}$) and stellar mass ($\Sigma_{\rm M}$) surface densities quite large for the knot 5.1, Log$_{10}(\Sigma_{\rm SFR}) \simeq 3.7$ M$_{\odot}$~yr$^{-1}$~kpc$^{-2}$ and Log$_{10}(\Sigma_{\rm M}) \simeq 4.1$ M$_{\odot}$~pc$^{-2}$, consistent with the values observed in local YMCs \citep[e.g.,][]{bastian06, ostlin07}.
A constant star formation history would imply a SFR of $\sim 3.3$ M$_{\odot}$~yr$^{-1}$ and the sSFR $\simeq 330$ Gyr$^{-1}$. Interestingly, such large densities and sSFR are lower limits since the SF history was likely not constant with time.
The presence of such YMC  influences the overall observed appearance (morphology and photometry) of the host galaxy (see Section \ref{sec:sunburst}); in this case, the star cluster is well recognized only because of strong gravitational lensing that boosts the observed flux and significantly increases the spatial resolution. 

\subsubsection{Escape velocity}
\label{escape}

Young massive star clusters are powerful producers of ionizing radiation, with an enormous sSFR and forging of hot and massive stars that demonstrate they can provide sufficient stellar feedback in such an early phase to perforate the interstellar medium of the galaxy. Observations of local young massive clusters indicate that even massive systems (M~$\simeq 10^{5}-10^{6}$ \msun) appear essentially gas-free even at very young ages (a few Myr, \citealt{cabrera15}), indicating that strong feedback processes must already be at play  in the earliest phases, in some cases before the onset of SN explosions. Again, the YMC hosted in the {\rm Sunburst} galaxy and the low-column-density of neutral gas along the line of sight (N$_{\rm HI} < 10^{17.2}$~cm$^{-2}$ given  by a large Lyman continuum escape fraction) may represent such a phase of intense stellar feedback caught in the act at 3 Myr since the initial burst of star formation. 

We estimate the escape velocity $v_{\rm esc}$ of the YMC 5.1 from the following equation \citep[][]{cabrera16}:

\begin{equation}
    v_{\rm esc} = f_{c} \sqrt{\frac{M_{\rm cl}}{R_{\rm eff}}} ~~ {\rm km~s^{-1}}
\label{eq:vesc}
,\end{equation}

\noindent where $M_{cl}$ is the cluster mass, $R_{\rm eff}$ the effective radius, and $f_c$ accounts for the dependence of the escape velocity on the density profile of the cluster. The value of $f_c$ ranges between the minimum and maximum, namely, 0.076 - 0.130 \citep[from Table~2 of][]{georgiev09}. Adopting M$_{\rm cl} = 10^7$ \msun\ and R$_{\rm eff}=8$ pc as determined above and the highest value of $f_c = 0.130,$ we get an upper limit to the escape velocity of the cluster, $v_{\rm esc} \simeq 145$ \kms.
The clear asymmetric blue tail of the \oiiiv\ line (shown in Figure~\ref{OIIIline}) implies an outflow component along the line of sight.
Following \citet{perna15}, we derive (1) the maximum velocity defined as the velocity at 2\% ($v02$) of the cumulative flux, $F(\rm V) = \int_{a}^{b} F_{v}(v') dv'$ (where F$_{\rm V}$ is the line profile in the velocity space and the position of $v=0$ of the cumulative flux is set at the systemic redshift given by the narrow component, $\lambda=16878.6$\AA, or z=2.3702, adopting 5008.24\AA\ \oiiialone\ vacuum wavelength) and (2) the line width w80, that is, the width comprising 80\% of the flux defined as the difference between the velocity at 90\% ($v90$) and 10\% ($v10$) of the cumulative flux of the entire line profile. 
The maximum velocity and the w80 are 380 and 260 \kms, which are also consistent with the
FWHM of the broad component of the double Gaussian fit mentioned above, 320 \kms. 
Overall, such outflow velocities are consistent with what is inferred by \citet{rivera17} on the same object based on absorption lines of silicon and eventually higher than the escape velocity of the cluster inferred above.
This also agrees with the scenario in which the most plausible source for driving the escape of ionizing radiation is stellar wind and radiation from hot massive stars \citep{izotov18_O32,heckman11}, which are certainly hosted in the cluster core \citep[][]{vanz_sunburst}.

\section{The {\rm Sunburst} galaxy}
\label{sec:sunburst}

In this section, we investigate the physical properties and the appearance of the {\rm Sunburst} galaxy on the source plane in the context of the aforementioned presence of star clusters. In particular, we emphasize the significant contribution of the YMC 5.1 to the optical nebular emission and the ultraviolet appearance of the entire galaxy. 

\subsection{Physical properties}
\label{sect:SEDandMORPH}

\begin{figure*}
        \centering
        \includegraphics[width=\linewidth]{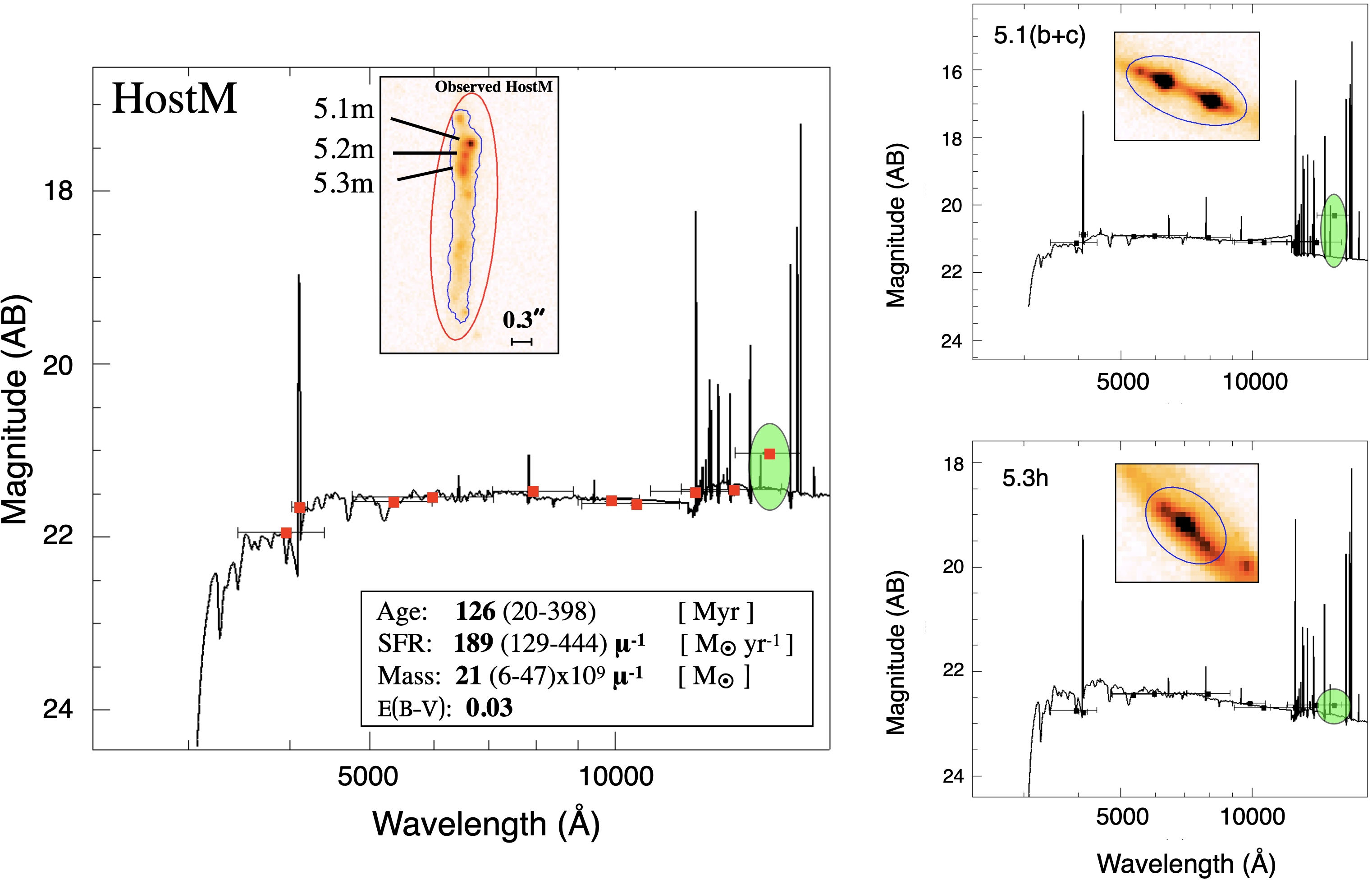}
        \caption{Nebular SED fitting of the {\rm Sunburst} galaxy, referred to as HostM in the text is shown in the main left panel. It is moderately magnified and represents the emission from the entire galaxy. In the inset, the F555W-band image shows HostM with the elliptical aperture (red line) used to compute PSF-matched photometry and the knots 5.1, 5.2 and 5.3. 
        The shaded green ellipses mark the photometric excess in the F160W band, due to nebular emission of \oiiiiv\, mainly due to the YMC 5.1. High S/N nebular SED-fit of YMC 5.1 is shown in the upper right panel, including both 5.1b and 5.1c to highlight the clear photometric jump in the F160W band, as confirmed from VLT/X-Shooter observations (Appendix~\ref{ha}).  The inset shows the images of 5.1b,c and the adopted aperture in the F555W band (blue).  Same SED-fit of the complex 5.3h is shown in the bottom right panel as an example in which no evident nebular contribution is present in the F160W band. The inset shows the image of 5.3h and the adopted aperture in the F555W band (blue ellipse).}
        \label{sed}
\end{figure*}

SED-fitting has been performed on the image ``m'' of the galaxy (dubbed {\it HostM} hereafter), which offers the closest view of the entire {\rm Sunburst} galaxy (see Figure~\ref{pano}, arc III, image "m"). This is indeed the least magnified image of the target and therefore the one that resembles more closely the intrinsic shape and morphology of the entire galaxy.
We use the HST images F390W, F410W, F555W, F606W, F814W, F098W, F105W, F140W, and
F160W available in the HST archive (see Sect.~\ref{data}). In order to obtain unbiased photometric measurements, we first estimated the point spread function for all the images using bright, unsaturated stars and then we PSF-matched all the bands to the F160W image using relevant convolution kernels.
Photometry of the HostM is extracted in the elliptical aperture shown in Figure~\ref{sed} using the software A-PHOT \citep[][]{merlin19}. The median(mean) magnification $\mu_{tot}$ within such an aperture is nearly constant, $19.1(19.3)$, with a standard deviation of 2.1. 
The SED-fitting on the multi-band photometry has been computed with the {\tt zphot.exe} code \citep{fontana2000}, as described in \citet{castellano16}
(see also \citealt{vanz19}) and the results are shown in Figure~\ref{sed}, while
the delensed physical quantities are reported in the last row of Table~\ref{knots}. \citet{Bruzual_2003} templates have been adopted with the following priors: exponentially declining star-formation histories with e-folding times of $0.1 \leq \tau \leq 15.0$ Gyr, a \citet{Salpeter_1955} initial mass function, and the extinction laws from both  \citet{Calzetti_2000} and Small Magellanic Cloud \citep{prevot84}. We considered the following range of physical parameters: $0.0 \leq E(B-V) \leq 1.1$,  Age~$>$~20Myr (defined as the onset of the star-formation episode) and metallicity Z/Z$_{\odot}$ = 0.02, 0.2, 1.0.  The resulting delensed ultraviolet magnitude at 1600\AA, stellar mass and SFR are 24.8, $\sim 1.1\times 10^{9}$ \msun, $\sim 10$ M$_{\odot}$~yr$^{-1}$, respectively, with an estimated age of $\simeq 130$ Myr and E(B-V) in the range $0.03-0.1$ (uncertainties are reported in the same figure). 
An additional clear photometric signature is the apparent
boost of flux in the F160W-band, indicated in Figure~\ref{sed}. 
The zooming provided by strong lensing on arcs I and II coupled with VLT/X-Shooter observations allow us to confirm that the origin of the photometric break is mainly due to knot 5.1 (i.e., the 3 Myr old YMC discussed above). Indeed, VLT/X-Shooter spectroscopy shows the presence of nebular lines, \hb\ and \oiiidoub\ emerging prominently from 5.1 (see Appendix~\ref{ha}).
In particular, as a test case, we compared the F160W photometric excess measured on the YMC (5.1b + 5.1c, Figure~\ref{sed}) and the corresponding line fluxes secured with VLT/X-Shooter on the same pair. If interpreted as a boost from the nebular line group 
\hb\ + \oiiidoub\ (magnitude 20.25 on the F160W band and 21.10 at continuum), such photometric jump corresponds to $(2.00\pm0.05)\times10^{-18}$~erg~s$^{-1}$~cm$^{-2}$~\AA$^{-1}$~\footnote{Calculated as $(c \lambda^{-2}) \times 10^{-0.4(m+48.59)}$, where $c$ is the speed of light, $\lambda$ the effective wavelength of the F160W-band (15405.2\AA) and $m$ the AB magnitude.}, which is fully consistent with what is inferred from the spectrum of the same pair of images, $(1.97\pm0.10)\times10^{-18}$~erg~s$^{-1}$~cm$^{-2}$~\AA$^{-1}$ (such values also correspond to a rest-frame equivalent width of $\simeq 1100$\AA\ for the group of lines). We then compared what contribution the YMC provides to the F160W excess of the whole galaxy HostM. To this aim, we used A-PHOT to extract photometry on image 5.1m within a circular aperture of $0.4''$ diameter ($\simeq 2 \times$ FWHM) and from the full image HostM (elliptical aperture, Figure~\ref{sed}), adopting the F140W band as a reference probe of the underlying continuum. It turns out that the YMC contributes to about 30\% of the observed excess on HostM. It is worth noting that its stellar mass accounts for only less than 1\% of the total stellar mass of the galaxy.

Other regions of the same galaxy, like 5.3, do not show such an excess (Figure~\ref{sed}). The overall SED of HostM includes the contribution of different star forming regions, with 5.1 showing the signature of a recent burst, whereas the rest of the galaxy might have more evolved characteristic features (e.g., relatively evolved star-forming regions with ages older than 3 Myr). 

\subsection{Morphology}

The delensed image of the galaxy HostM is shown Figure~\ref{HostM}. This reconstruction is obtained as follows. We start with one of the multiple images of the Sunburst arc, namely: the arc segment shown in inset III of Figure~\ref{pano} (also displayed in Figure~\ref{HostM}). We super-sampled the arc on a regular grid whose pixel scale is $0.01$ arcsec and we use the ray-tracing algorithm implemented in the software {\tt SkyLens} \citep{Meneghetti_2008,Meneghetti_2010,Meneghetti_2017,Plazas_2019} to map the surface brightness in each pixel onto the source plane. 
We chose to super-sample the image at a resolution of $0.01$ arcsec/pixel in order to increase the number of support points on the source plane where we mapped the image surface brightness. The resulting unstructured data points are then interpolated on another regular grid, the resolution of which is again $0.01$ arcsec. We convolved the image with a model of the HST PSF in the F555W band, obtained using the software Tiny Tim \citep{krist_2011}. Finally, we re-sampled the image at the resolution of $0.03$ arcsec per pixel.     
\footnote{The observed image of the arc used to reconstruct the galaxy HostM is already convolved with the HST PSF. When re-convolving the unlensed image,  our procedure does not account for this pre-existing convolution. By de-lensing the arc, its size, including that of the PSF, is reduced by a factor $\mu_{tot} \simeq 19$. Thus, the resulting PSF size on the source plane is so small that we can safely neglect it.}

As expected, without strong lensing (i.e., assuming $\mu=1$), the HostM would appear
as a galaxy sampled by $\lesssim4$ HST resolution elements (FWHM~$=0.1''$, in the F555W band, Figure~\ref{pano}). The same figure also shows that the YMC (5.1) and the two 5.2 and 5.3 clusters merge into a nucleated unresolved component, which would be recognized as a $\sim 1$ kpc-size star-forming clump (adopting the $0.1''$ PSF in the UV, or 250 pc per pixel $0.03''$). Future facilities assisted by extreme adaptive optics working at a resolution of a few tens of milliarcsec will be capable of discerning these structures also at the magnification of HostM (but see Sect.~\ref{future}). In the framework of the LyC galaxies, the observation in unlensed fields with medium-resolution spectroscopy ($R>4000$) of sources leaking LyC might capture ionized channels possibly associated to underlying (and totally unresolved) star clusters. In fact, in the case of {\rm Sunburst} the detection of a prominent \lya\ emission at the systemic redshift for the YMC \citep[][]{rivera17}, in addition to a more regular and typical broader emission generated by radiative transfer effects, is consistent with the scenario in which an ionized channel was carved along the line of sight (corroborated by the LyC detection presented in \citealt{rivera19}, and as predicted by \citealt{behrens14}).
In unlensed fields, such \lya\ emission at systemic redshift has been identified for other LyC galaxies showing nucleated ultraviolet morphology; this, if compared to {\rm Sunburst} \citep[][]{vanz_sunburst}, might imply young massive star clusters are acting as prodigious ionizing photon producers 
and drillers within the interstellar medium. It is still not clear if this LyC escaping radiation mode is typical at redshift $2-3$ \citep[][and references therein]{matthee21} or even during reionization.

\section{High cluster formation efficiency in the {\rm Sunburst} galaxy}
\label{sec:CFE}

As discussed in Sect.~\ref{clusters}, we identified at least 13 likely star clusters (see Table~\ref{knots}). Taking advantage of the new lens model \citep{pignataro21}, we can compare the integrated ultraviolet light of all star clusters to the luminosity of the galaxy HostM (known as T$_{\rm L}$(UV) parameter). We follow two
ways of calculating such a fraction of light: (a) by comparing the delensed ultraviolet luminosity of the clusters and the host galaxy (dependent on the lens model) and (b) by comparing the ultraviolet light self consistently within the observed image HostM, independently from the lens model (this is the conservative approach). 
In case (a) the direct sum of delensed fluxes of the star clusters reported in Table~\ref{knots} accounts for 55\% of the UV light of HostM (M$_{1600}=-20.3$). 
The uncertainly on T$_{\rm L}$(UV) derived in this way is at least 50\% of its value and is dominated by errors on the magnification factors (arcs I and II). In case (b) the fraction of the UV light is measured from the image HostM, which is nearly uniformly and moderately amplified (median $\mu_{tot}=19$). In this case, a relative measure of the UV light can be computed independently from magnification. 

\begin{figure*}
        \centering
        \includegraphics[width=\linewidth]{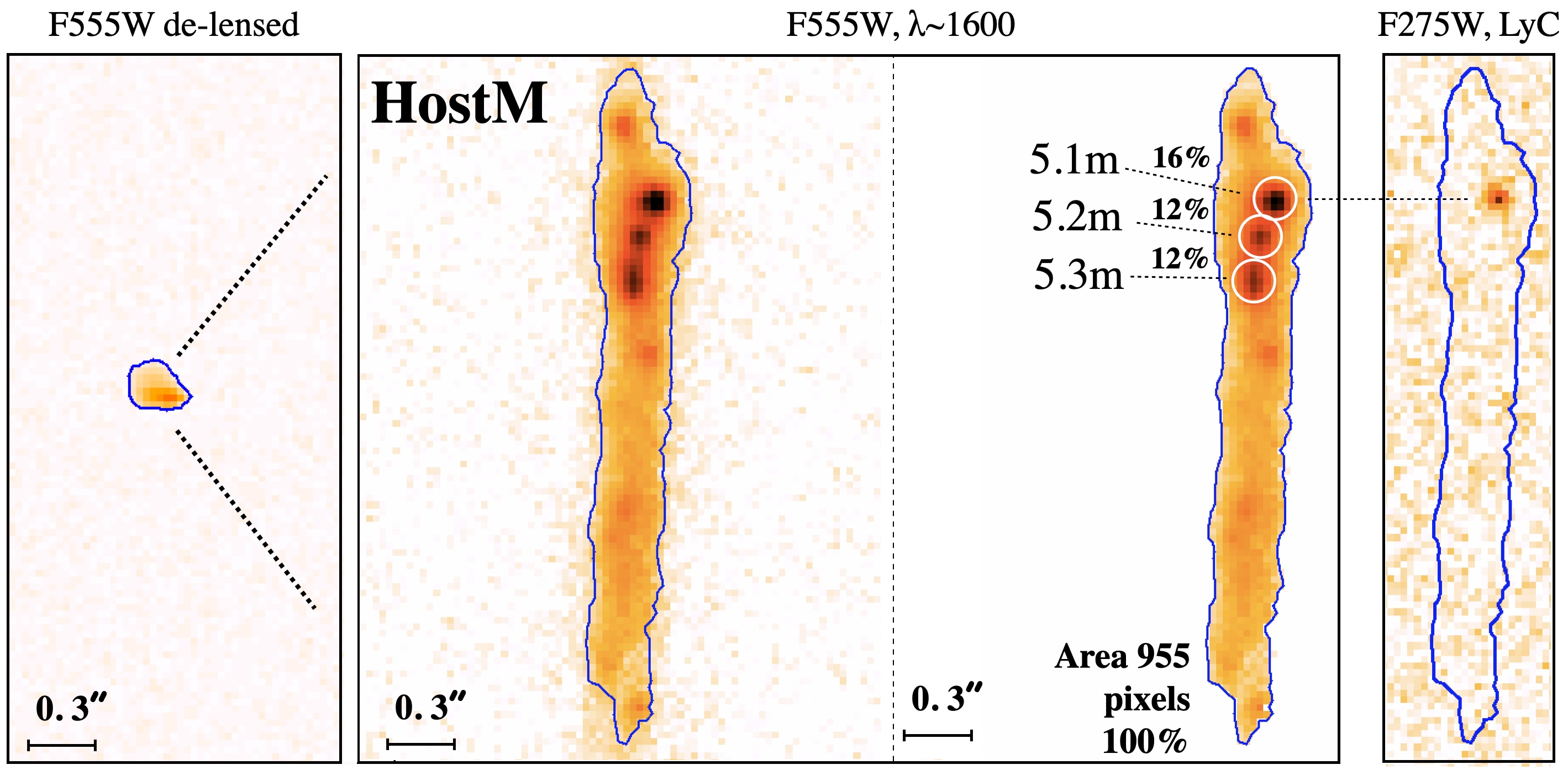}
        \caption{ Reconstruction of the HostM galaxy in the source plane (delensed) is shown in the left panel, with almost all the details merged into a compact, barely resolved source. {\it Middle panels}:  Galaxy HostM in the F555W-band and the segmentation image with the blue contour at 5$\sigma$ (left and right, respectively). The three regions 5.1, 5.2, and 5.3 (white apertures with $0.2''$ diameter) containing star clusters account for at least 40\% of the ultraviolet light of the galaxy. {\it Right panel}: Same region of the galaxy is outlined with the blue contour in the F275W band probing LyC. It is clear the conspicuous escape of ionizing radiation from the YMC 5.1.}
        \label{HostM}
\end{figure*}
The ultraviolet flux of HostM shown in Figure~\ref{HostM} is compared to the combined flux emerging from 5.1m, 5.2m and 5.3m, calculated within circular apertures of diameter $0.2''$ ($\simeq 2 \times$~FWHM of the F555W band).
The resulting T$_{\rm L}$(UV) from 5.1, 5.2, and 5.3 is at least $\sim 40\%$ of the total 1600\AA\ ultraviolet light inferred from the host galaxy, while the single YMC 5.1 accounts for 16\%. These values should be regarded as lower limits, since the other star clusters are not included in the calculation. Therefore the two inferred T$_{\rm L}$(UV) from case (a) and (b) are consistent, suggesting that  T$_{\rm L}$(UV) spans the range between 40-60\%.
The fact that star clusters significantly contribute to the ultraviolet light might also suggest that a significant fraction of the star formation in the galaxy efficiently occurred in such objects. 

Indeed, the cluster formation efficiency ($\Gamma$) 
is effectively a measurement of the total stellar mass forming in bound star clusters with respect to the total stellar mass forming in the galaxy, referring to the same interval of time \citep[][]{bastian08}.
Following \citet{adamo15} $\Gamma$ is defined as the star cluster formation rate (CFR) over the star formation rate of the host galaxy, both referred to the same interval of time, {\it dt}. The total stellar mass formed in clusters during {\it dt} is the CFR.
In the Local Universe ($\lesssim 20$ Mpc distance), the CFR is typically estimated by integrating the stellar mass in bound clusters younger than 10 Myr ({\it dt}) above an established mass limit. The undetected part of the mass distribution is included by extrapolating over an assumed (and properly normalized) star cluster mass function down to 100 \msun\ \citep[][]{adamo17}. Giving the complexity introduced by the lens we do not correct for such a mass limit, therefore, the inferred $\Gamma$ can be considered a lower limit (such a limit does not alter the conclusion of this section).
The age range used to derive the CFR is limited  by  the time  scales  over which the available SFR tracer are sensitive to. 
Finally, the resulting $\Gamma$ is simply the ratio between the CFR and the SFR of the galaxy, expressed on a comparable time scale.

At high redshift such calculation is extremely challenging for two reasons:
(I) gravitationally bound star clusters need to be identified and (II) the formation time scale for both clusters and the hosting galaxy need to be understood. 
The first point is partially addressed thanks to lensing amplification (Sect.~\ref{clusters}), the second is the most uncertain since star formation rate indicators on short timescales ($\lesssim 10$ Myr) would be needed for each knot (e.g., \ha, see Sect.~\ref{future}).
The integrated stellar mass of bound star clusters reported in Table~\ref{knots} is $\gtrsim 70\times 10^{6}$ \msun, under the aforementioned assumptions of instantaneous burst and age of 3 Myr for the knot 5.1 \citep[][]{chisholm19} and 7 Myr for the others (Appendix~\ref{ha}). 
The time interval within which such a fraction of stellar mass formed cannot be constrained from the available data. However, it is worth noting that if we relax the ages of the clusters to 50 Myr (fixing the age of the knot 5.1 to 3 Myr), the integrated mass would 
exceed the total stellar mass of the galaxy derived from the SED fitting ($\simeq 10^{9}$\msun). Although this argument might appear stuck in a vicious circle, it might nonetheless suggest the sample of 13 clusters discussed in this work was likely formed in the recent past of the galaxy ($<50$ Myr). 
Therefore, adopting a formation timescale of the star clusters of 10(20)(50) Myr, we have $\Gamma_{10,20,50} > 60(30)(12)$\%. 
The lower limit is due to the fact that forming star clusters fainter than the detection limit are not considered. As mentioned above, the complex geometry of the lens currently prevents us from correcting for such an incompleteness and will require a second version of the lens model (e.g., by adding more multiple images or constraints from galaxies at different redshifts), so we rely on what is detected and consider it a lower limit (a future analysis will take into account this lack of completeness).
Adopting a fiducial time scale of 20 Myr for the CFR and SFR $\simeq 10$ \sfr\ for the HostM, then the cluster formation efficiency $\Gamma$ would exceed 30\%. Both $\Gamma$ and T$_{\rm L}$(UV) suggest there was a vigorous formation of star clusters in the {\rm Sunburst} galaxy, in which a significant fraction of the stellar mass was produced in star clusters.

\begin{figure}
        \centering
        \includegraphics[width=\linewidth]{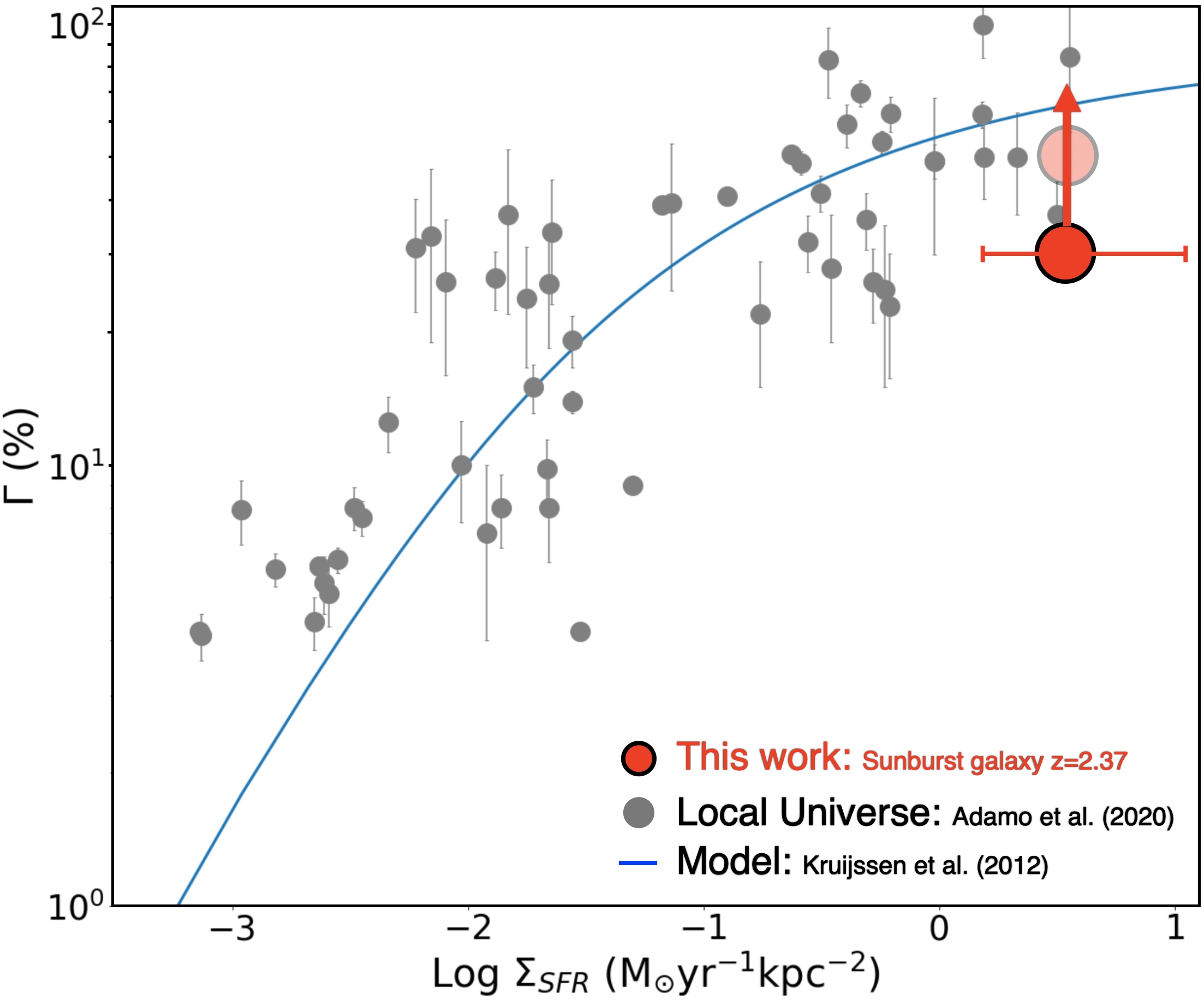}
        \caption{Cluster formation efficiency, $\Gamma,$ as a function of the star-formation rate surface density $\Sigma_{\rm SFR}$. The compilation of gray symbols represent estimates in the local Universe taken from \citet{adamo20,adamo20extreme}. The solid blue line reproduces the \citet{kruijssen12} fiducial model. 
        The estimated lower limit on $\Gamma$ (calculated over 20 Myr time scale, see text for details) for the {\rm Sunburst} galaxy is reported with the filled red circle at 30\%, while the transparent circle is the estimate fraction of the ultraviolet light (T$_{\rm L}$(UV)) computed from the delensed UV light integrated over all the star clusters reported in Table~\ref{knots}. The inferred quantities are likely lower limits, as additional clusters might have remained undetected.}
        \label{gamma}
\end{figure}

It is worth noting that such a high value of $\Gamma$ is consistent with what is observed in the Local Universe. The observed area of HostM shown in Figure~\ref{HostM} corresponds to $\simeq 3$ sq.kpc on the source plane, which implies a relatively high Log$_{10}(\Sigma_{\rm SFR}) \simeq 0.5_{-0.2}^{+0.3}$ M$_\odot$ yr$^{-1}$ kpc$^{-2}$ (independent from magnification and with the 68\% confidence interval on the SFR from SED fitting). {\rm The Sunburst} galaxy falls in the upper right part of the $\Gamma-\Sigma_{\rm SFR}$ relation, mimicking the local extreme forming galaxies described by \citet{adamo20extreme} (see Figure~\ref{gamma}). Interestingly, this is a region where similar local LyC leakers have also been identified \citep[][]{keenan17haro11, ostlin2haro11_LyC}.

\section{Escaping Lyman continuum from the galaxy}
\label{sec:fesc}

\citet{rivera19} estimated the escape fraction, {\tt fesc}, of the 12 multiple images of 5.1(a$-$n), assuming an intergalactic attenuation and {\tt Starburst99} models \citep[][]{leitherer14}.
Following the standard formalism for the estimation of the relative escape fraction \citep[e.g.,][]{steidel2001}, {\tt fesc,rel}, we have for 5.1:
\begin{equation}
    {\tt fesc,rel(5.1)} = \frac{F(F814W)/F(275W)_{intr}}{F(F814W)/F(275W)_{obs}} \times \frac{1}{T(IGM)}
    \label{eq:fesc}
,\end{equation}

\noindent where T(IGM) is the intergalactic attenuation of the LyC probed with the F275W band (T(IGM)~$=1$ meaning no attenuation) and the flux density ratios, intrinsic ("intr") and the observed ("obs"), at the given positions $a-n$.
The estimated escape fraction quantity has been observed to vary among the multiple images of 5.1, as it is likely modulated by the differential amount of \hi\ gas along the different paths. Overall, \citet{rivera19} reported a line-of-sight {\tt fesc,rel}~$=93^{+7}_{-11}$\% with 46\% as a robust lower limit.

The next step is to estimate the (observed) {\tt fesc,rel} associated to the entire galaxy. We can adopt the same quantities presented above and include in Eq.~\ref{eq:fesc} the observed flux density ratio referred to the whole galaxy observed in the segment, $m$, which has an observed magnitude: F814W = 21.60 (which corresponds to 24.8 delensed, Figure~\ref{sed}). The measured F275W and F814W magnitudes of the knot 5.1m are 27.58 (at S/N=6.2) and 23.85 (at S/N$>$40) 
\citep[Table~1 of][]{rivera19}.
The magnitude contrast in the F814W-band between the YMC (5.1m, with observed F814W=23.85) and the host galaxy (HostM, with observed F814W=21.60) implies a reduction of {\tt fesc,rel} by a factor 8. Therefore {\tt fesc,rel} of the entire galaxy is $\sim 12$\%, with a lower limit of 6\%. Such values should further be considered lower limits since there might be missing LyC signal emerging from other regions of the galaxy which are fainter than the current F275W depth. 
It is worth noting that the YMC has an intrinsic LyC and 1600\AA~magnitudes of 30.6 and 26.9, respectively, after correcting for magnification at the predicted position of 5.1m ($\mu_{tot} \simeq 16.1$).
Such values confirm the difficulty in securing relatively bright (sub-L$^{\star}$) high redshift LyC galaxies in unlensed fields down to {\tt fesc,rel}~$\sim10$\%.
Those currently confirmed at high redshift are relatively bright LyC leakers and might be the result of hidden and multiple (spatially indistinguishable) star clusters
\citep[e.g.,][]{steidel18,pahl21_LyC_clean,vanz_ion2,vanz18,barros16,vanz_sunburst}. In the case of {\rm Sunburst} galaxy, a single YMC dominates the observed LyC leakage.
However, it is worth noting that in general multiple and coeval YMCs containing massive O-type stars might concur to produce an integrated large ionizing photon production efficiency, observable through prodigiously large photometric excess due to nebular lines in the optical rest-frame bands, much stronger than what is reported here (see Sect.~\ref{CFE_EOR}). 
In this regard, it is worth referring to the recently discovered super-bright unlensed LyC galaxy by \citet{RUI2021} showing QSO-like luminosity, M$_{\rm UV} = -24.11$, large sSFR~$\simeq 100$ Gyr$^{-1}$, and Log$_{10}(\Sigma_{\rm SFR})\sim 2.2$, implying a large stellar cluster formation efficiency ($\Gamma$) according to the relation shown in Figure~\ref{gamma}.

\section{Summary and conclusions}
\label{sec:summary}

The {\rm Sunburst} galaxy is an exceptional laboratory for investigating high-$z$ star formation for two main reasons: (1) individual forming stellar clusters can be identified at cosmological distance and (2) the link between the LyC leakage of the galaxy and the presence of such young (and bursty) star clusters can be seen. The main results of this study are reported in the following:

\begin{enumerate}

\item{{\it Star clusters at cosmological distance}. The {\rm Sunburst} arcs are the result 
of a large lensing tangential amplification of a dwarf galaxy, in which star-forming regions down to a few parsec scale are probed. More than 50 multiple images of at least 13 star-forming knots belonging to the same galaxy have been identified (a panoramic view is shown in Figure~\ref{pano}). 
In particular, taking advantage of a new accurate lens model \citep{pignataro21}, we find that such knots are likely gravitationally bound stellar clusters with stellar masses and effective radii spanning the range $\simeq 10^{5-7}$ \msun\ and $\simeq 1-20$ pc. The resulting cluster formation efficiency of the galaxy is $\Gamma > 30\%$, placing it in the regime of extreme physical conditions when compared to local galaxies showing similar large $\Gamma$ \citep[][]{adamo20extreme}.}

\item{{\it The region emitting LyC is a young massive star cluster}. The knot 5.1 is the youngest (3 Myr, \citealt{chisholm19}), the most massive ($10^{7}$ \msun\ both from photometric and dynamical estimates) and the brightest (M$_{\rm UV} = -18.6$) among the identified star clusters. Such a YMC shows an effective radius smaller than 10~pc and evident presence of hot massive (including Wolf-Rayet) stars (based on evident detection of P-Cygni profiles of \nv, \civmed\ and broad \heii, \citealt{chisholm19,vanz_sunburst}); LyC emerges from this cluster with large relative escape fraction of $43-93$\%, depending on which multiple image is used for the calculation \citep[][]{rivera19}.
The delensed LyC(1600\AA) magnitude is 30.6(26.9).
The sSFR is very large $>330$ ~Gyr$^{-1}$, consistently with the prominent optical nebular emissions lines, e.g., \hb\ + \oiiidoub\ rest-frame equivalent width of $\gtrsim 1000$~\AA\ (as inferred from VLT/X-Shooter spectroscopy) and as expected in such bursty events exhibiting ionizing photon production efficiency that is greater than the canonical values \citep[][]{chevallard18}. The O32 index (\oiiidoub / \oiidoub) estimated on the YMC is $17 \pm 3$, consistent with the necessary condition of having a LyC leakage (\citealt{barrow20}, see also \citealt{izotov18_O32}).}

\item{{\it The hosting galaxy}. {\rm Sunburst} would appear as a relatively small ($\simeq 3$ sq.kpc) and low mass ($\simeq 10^{9}$~\msun) galaxy  with a dense star formation rate surface density (Log$_{10}(\Sigma_{\rm SFR})\simeq 0.5$) and large sSFR~$\gtrsim 10 $~Gyr$^{-1}$. It would be marginally resolved in the UV, nucleated and dominated by the YMC 5.1 SF knot UV emission, together with 5.2 and 5.3, which would be recognized as a kpc scale SF clump without lensing. The {\tt fesc,rel} of the galaxy is $>6-12\%$.
The lower limit is due to the fact that additional $-$ possibly missing $-$ LyC radiation might escape along the l.o.s. from other regions of the galaxy, but not probed  with current depth yet. The photometric jump observed in the F160W-band (relative to the continuum probed in the F140W-band) implies a rest-frame 
equivalent width of the line complex \hb\ + \oiiidoub\ of $\simeq 450$\AA, of which $\simeq 30$\% of the signal is due to the single YMC 5.1. It is worth noting that the same YMC 5.1 represents  $\lesssim 1$\% of the total stellar mass of the galaxy.}

\end{enumerate}

\subsection{Plausible indirect evidence of a high cluster formation efficiency at high redshift (and during reionization) }
\label{CFE_EOR}

Studies in the local Universe have already demonstrated the key role of super star clusters as sources of intense stellar feedback that significantly affects the interstellar medium of the host galaxies (ionization, cavities, or outflows, e.g., \citealt{james16, adamo20extreme}), eventually carving ionized optically thin channels to LyC photons \citep[e.g.,][]{micheva17,bik18}. The aforementioned properties of the {\rm Sunburst} galaxy resemble those of confirmed unlensed LyC leakers at $z\lesssim4$ \citep[e.g.,][]{vanz_ion2, vanz18, barros16, schaerer16, izotov16Nature, izotov21lowmass,RUI2021} and those of $z\sim 2-3$ analogs of $z>6.5$ sources contributing to cosmic reionization \citep[][]{du20_z2_analogs_toEOR, mengtao21,katz20,matthee21}.
For example, the photometric discontinuity observed with Spitzer/IRAC at $z \sim 6.5-8$ due to optical nebular emission lines is also a distinctive signature present in the {\rm Sunburst} SED for the same rest-optical lines \citep[e.g.,][]{castellano17,smit14neb,smit15neb,smit16neb,borsani16,laporte14,finkelstein13nature,barros2019}. 
In particular, \citet{endsley21} find 
a large median (\hb\ + \oiiialone) EW of $760\pm100$\AA, with a significant fraction ($23\pm7$\%) showing extreme values larger than 1200\AA. Similarly, \citet{castellano17} derive EW of $1500\pm 500$\AA\ at $z\simeq 7$ from Spitzer/IRAC color excess, as well as \citet{barros2019} at $z\simeq 8$. 
Such results imply a high specific star formation rate (sSFR~$>10-35$~Gyr$^{-1}$) and suggest intense bursts are common in these sources during reionization. 
Conversely, the occurrence of strong (\hb~+~\oiii) emission is only a few percent at lower redshift, $z\simeq 2-3$ \citep[e.g., Figure~9 of][and references therein]{du20_z2_analogs_toEOR}, suggesting that the frequency of such
phases of intense star formation evolves with cosmic time, which appears to cycle regularly in the reionization era \citep[][]{endsley21}. 
A possible explanation is that the net increase of the SFR surface density with increasing redshift (up to $z\sim 8$, \citealt{naidu20}) would imply a high cluster formation efficiency \citep[as observed for dense star-forming galaxies in the local Universe,][]{adamo20extreme},
namely: a prodigious ``burstiness'' due to actively forming star clusters. An increase of $\Gamma$ with redshift has also been calculated by \citet{pfeffer18} in cosmological simulations that follow the co-evolution of galaxies and their star cluster populations, assuming the observed local $\Gamma - \Sigma_{\rm SFR}$ relation. 

The {\rm Sunburst} galaxy described in this work represents the first example to demonstrate that prominent optical-rest emission lines, a high cluster formation efficiency $\Gamma > 30-50$\%, relatively large sSFR~$\sim 10$~Gyr$^{-1}$, 
and Lyman continuum leakage can co-exist together in the same system. From the observational point of view, such objects as {\rm Sunburst} represent the ``Rosetta stone'' of stellar ionization, which might be rare at $z \sim 2$, but possibly more frequent at $z>6.5$ \citep[e.g.,][]{matthee21}. 
The presence of older (than 3 Myr) massive star clusters in the galaxy suggests that similar (bursty) conditions where in place in the past, potentially favoring an intermittent behavior of the escaping LyC radiation and ionizing production \citep[][]{trebitsch2017,wise2014}.
{\rm Sunburst} is not as extreme as other LyC sources observed at $z\simeq 3$ or in the local Universe, for which (\hb\ + \oiiialone) EW exceeds 1000-2000\AA\ rest-frame (e.g., at z=3.2 \citealt{vanz_ion2,barros16} or in the Local Universe, e.g., \citealt{izotov21lowmass}). 

It is worth noting that young star clusters are the sites where: (1) the majority ($>90\%$) of ionizing massive stars (O-type) are forged \cite[][]{Vargas_Salazar20} and (2) stellar feedback consequently originated \citep[e.g.,][]{bik18}. 
Therefore, if star-formation drives reionization, then such aggregates are likely the key ionizing agents whose occurrence might be higher at high redshift when nebular emission is observed to be most prominent \citep[][]{endsley21,castellano17,barros2019}.

It is worth considering the possibility that the presence of multiple and almost coeval massive young star clusters might concur to elevate the ionizing photon production, especially if the truncation mass of the initial cluster mass function increases with redshift \citep[e.g.,][]{pfeffer18}. Such conditions might also increase the {\tt fesc,rel}, as multiple ionized channels can be carved in the ISM.
Local examples rich of young star clusters and high cluster formation efficiency are the dwarf starburst galaxies NGC 5253
\citep[][]{calzetti15_ngc5253}, NGC~3125 \citep[][]{wofford14}, or NGC~1569 \citep[][]{anders04}. Finally, it is worth noting that local LyC leakers show a rather nucleated morphology -- implying high SFR densities \citep[][]{izotov18_O32}, which in turn, follows from the findings of \citet{adamo20} as well as the details in Figure~\ref{gamma}; this would suggest young star clusters are present (though not spatially recognizable) and possibly playing a key role in carving ionizing channels, as in the case of {\rm Sunburst}.

\begin{figure*}
        \centering
        \includegraphics[width=\linewidth]{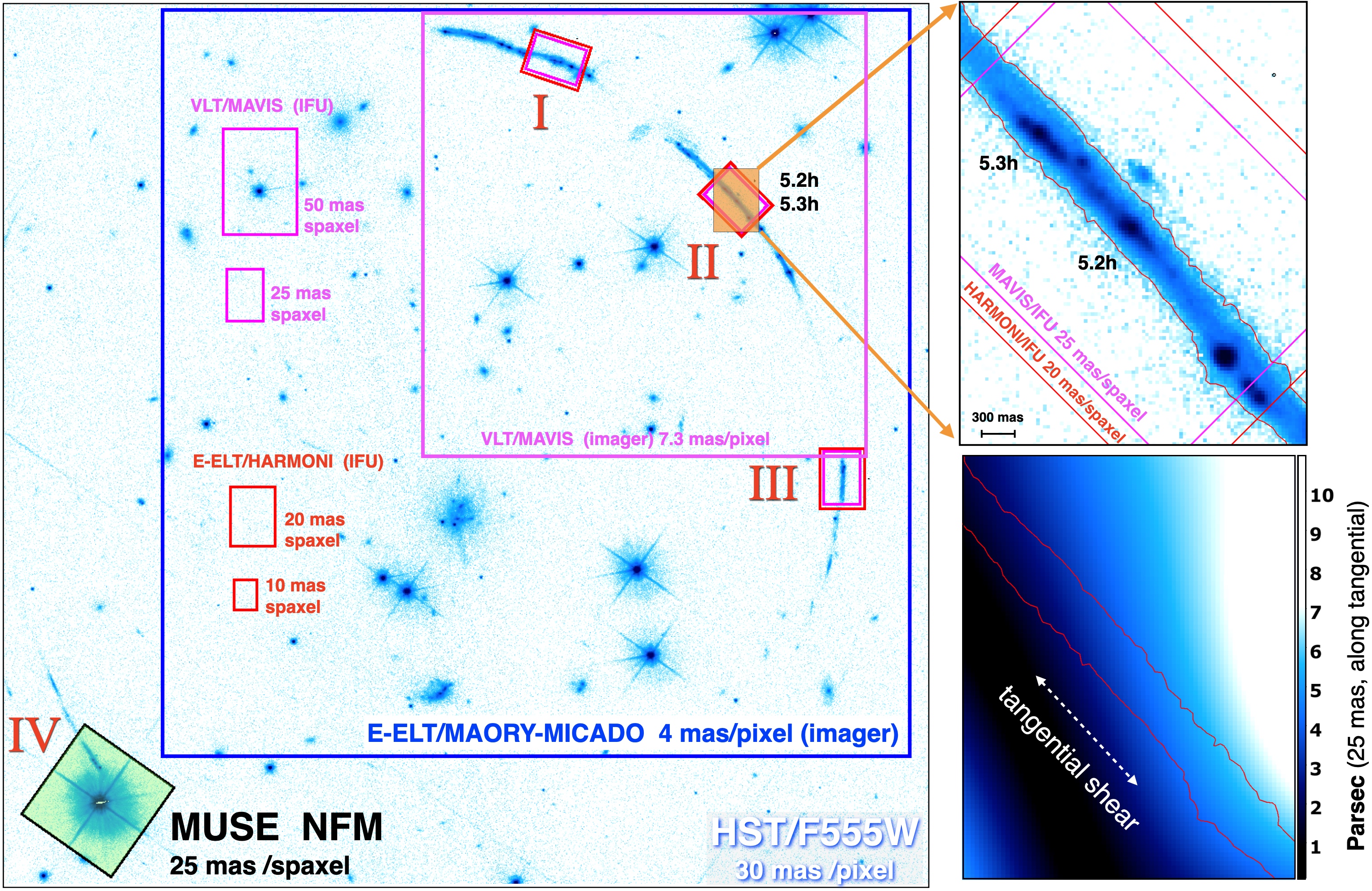}
        \caption{Main panel shows HST/F555W-band image (30 mas/pixel) with superimposed the field of views (FoV) of future instruments, such as VLT/MAVIS (IFU and imaging modes), ELT/MAORY-MICADO (imaging), and HARMONI (IFU), along with the layout of the MUSE-NFM targeting arc IV (transparent green box in the bottom-left corner, presented in Sect.~\ref{appendix_NFM}). The four arcs of {\rm Sunburst} are indicated with I, II, III, IV. MUSE-NFM requires the presence of a very bright star within $3.4''$ from the target, whilst future AO facilities will definitely probe stellar clusters down to the parsec scale in the most magnified regions. 
        The left panels show the zoomed region indicated with the orange shaded rectangle located in the main panel over the objects 5.2h and 5.3h (left). In particular, the region enclosing the knots 5.2h and 5.3h will be covered with MAVIS and HARMONI IFUs which will probe 1-5 parsec per spaxel, while imaging with MAVIS and MAORY-MICADO (wide blue and magenta squares in the main panel) will probe the parsec scale in the optical and near-infrared, respectively. 
    }  
        \label{fig:future}
\end{figure*}

\subsection{Current limitations and future prospects}
\label{future}

The presence of a bright star (H $<14$ in Vega mag) only  $3.3''$ away from arc IV of {\rm Sunburst} allowed us to use the VLT/MUSE narrow field mode (NFM) with optimal AO correction (see Figure~\ref{fig:future}). Such an XAO correction provides PSF in line with the expectations, even better than 60 mas. Details of these observations are presented in Appendix~\ref{appendix_NFM}.
There are two key aspects regarding MUSE-NFM worth mentioning:
the one hand, we found the achieved angular resolution was optimal (FWHM $\lesssim 60$ mas at $\lambda > 6000$\AA) despite the relatively high airmass ($\sim 1.7$), showing that MUSE-NFM equals HST imaging at similar wavelengths (see Figure~\ref{NFM}). A resolution of 60 mas on the counter-arc corresponds to $\sim 60$ parsec on the source plane along the tangential direction.

On the other hand, the very limited sky coverage offered by MUSE-NFM (e.g., the lack of a very close bright star) prevents us from targeting very magnified regions of the {\rm Sunburst} arcs, such arc II, in which a 60 mas resolution would have secured spectroscopy at 5-10 parsec scale. MUSE-WFM on that arc, assuming seeing $0.8''$ and $\mu_{tang}=50-100$ probes $131-65$ pc scale.

For this reason, an increased sky coverage in which XAO can be performed on (e.g., arc II) is key in such studies, allowing us to reach a spatial resolution of a few pc. In general, for super-lensed targets like {\rm Sunburst} \citep[e.g.,][]{sharon20, rigby18_II, rigby18_I}, VLT/MAVIS and ELT MAROY/MICADO or HARMONI are complementary (in the wavelength domain) and will provide a census of the cluster formation efficiency across cosmic epochs, their role in shaping the galaxy growth and ionizing output.
In this respect, ELT HARMONI will be ideal for probing the \ha\ of each SF knot in the near infrared wavelengths, while VLT/MAVIS will complement on similar angular resolution in the ultraviolet and optical bands.
The JWST integral field spectroscopy will also map \ha\ line emission, which will be free from sky lines and background.

\begin{acknowledgements}
      We thank the anonymous referee for the careful reading and constructive comments. 
      We thank A. Adamo for very stimulating discussion and providing the measures of $\Gamma$ in the local Universe (shown in Figure~\ref{gamma}).
      EV thank D. Calzetti for the precious clarifications on the formation channels of O-type stars. EV thank A. Renzini for illuminating discussions about any aspect of star, star-cluster and galaxy formation, and J. Chisholm for the stimulating interactions about LyC galaxies. We thank A. Fontana for making available the zphot.exe code.
      This project is partially funded by PRIN-MIUR 2017WSCC32 ``Zooming into dark matter and proto-galaxies with massive lensing clusters''. We acknowledge funding from the INAF 
      for ``interventi aggiuntivi a sostegno della ricerca di main-stream'' (1.05.01.86.31).  PB acknowledges financial support from ASI though the agreement ASI-INAF n. 2018-29-HH.0. MM acknowledges support from the Italian Space Agency (ASI) through contract ``Euclid - Phase D" and from the grant MIUR PRIN 2015 ”Cosmology and Fundamental Physics: illuminating the Dark Universe with Euclid”. FC acknowledges support from grant PRIN MIUR 2017 $-$ 20173ML3WW$\_$001. CG acknowledges support by VILLUM FONDEN Young Investigator Programme through grant no. 10123. KC acknowledges funding from the ERC through the award of the Consolidator Grant ID~681627-BUILDUP. 
      GBC acknowledges the Max Planck Society for financial support through the Max Planck Research Group for S. H. Suyu and the academic support from the German Centre for Cosmological Lensing. This research made use of the following open-source packages for Python and we are thankful to the developers of these: Matplotlib \citep{matplotlib2007}, MPDAF \citep{MPDAF2019}, PyMUSE \citep{PyMUSE2020}, Numpy \citep[][]{NUMPY2011}.
\end{acknowledgements}

\bibliographystyle{aa}
\bibliography{bibliography}

\begin{thebibliography}{119}
\expandafter\ifx\csname natexlab\endcsname\relax\def\natexlab#1{#1}\fi

\bibitem[{{Adamo} {et~al.}(2020{\natexlab{a}}){Adamo}, {Hollyhead}, {Messa},
  {Ryon}, {Bajaj}, {Runnholm}, {Aalto}, {Calzetti}, {Gallagher}, {Hayes},
  {Kruijssen}, {K{\"o}nig}, {Larsen}, {Melinder}, {Sabbi}, {Smith}, \&
  {{\"O}stlin}}]{adamo20extreme}
{Adamo}, A., {Hollyhead}, K., {Messa}, M., {et~al.} 2020{\natexlab{a}}, arXiv
  e-prints, arXiv:2008.12794

\bibitem[{{Adamo} {et~al.}(2015){Adamo}, {Kruijssen}, {Bastian}, {Silva-Villa},
  \& {Ryon}}]{adamo15}
{Adamo}, A., {Kruijssen}, J.~M.~D., {Bastian}, N., {Silva-Villa}, E., \&
  {Ryon}, J. 2015, \mnras, 452, 246

\bibitem[{{Adamo} {et~al.}(2017){Adamo}, {Ryon}, {Messa}, {Kim}, {Grasha},
  {Cook}, {Calzetti}, {Lee}, {Whitmore}, {Elmegreen}, {Ubeda}, {Smith},
  {Bright}, {Runnholm}, {Andrews}, {Fumagalli}, {Gouliermis}, {Kahre}, {Nair},
  {Thilker}, {Walterbos}, {Wofford}, {Aloisi}, {Ashworth}, {Brown}, {Chandar},
  {Christian}, {Cignoni}, {Clayton}, {Dale}, {de Mink}, {Dobbs}, {Elmegreen},
  {Evans}, {Gallagher}, {Grebel}, {Herrero}, {Hunter}, {Johnson}, {Kennicutt},
  {Krumholz}, {Lennon}, {Levay}, {Martin}, {Nota}, {{\"O}stlin}, {Pellerin},
  {Prieto}, {Regan}, {Sabbi}, {Sacchi}, {Schaerer}, {Schiminovich}, {Shabani},
  {Tosi}, {Van Dyk}, \& {Zackrisson}}]{adamo17}
{Adamo}, A., {Ryon}, J.~E., {Messa}, M., {et~al.} 2017, \apj, 841, 131

\bibitem[{{Adamo} {et~al.}(2020{\natexlab{b}}){Adamo}, {Zeidler}, {Kruijssen},
  {Chevance}, {Gieles}, {Calzetti}, {Charbonnel}, {Zinnecker}, \&
  {Krause}}]{adamo20}
{Adamo}, A., {Zeidler}, P., {Kruijssen}, J.~M.~D., {et~al.} 2020{\natexlab{b}},
  \ssr, 216, 69

\bibitem[{{Anders} {et~al.}(2004){Anders}, {de Grijs}, {Fritze-v. Alvensleben},
  \& {Bissantz}}]{anders04}
{Anders}, P., {de Grijs}, R., {Fritze-v. Alvensleben}, U., \& {Bissantz}, N.
  2004, \mnras, 347, 17

\bibitem[{{Bacon} {et~al.}(2012){Bacon}, {Accardo}, {Adjali}, {Anwand},
  {Bauer}, {Blaizot}, {Boudon}, {Brinchmann}, {Brotons}, {Caillier}, {Capoani},
  {Carollo}, {Comin}, {Contini}, {Cumani}, {Daguis}, {Deiries}, {Delabre},
  {Dreizler}, {Dubois}, {Dupieux}, {Dupuy}, {Emsellem}, {Fleischmann},
  {Fran{\c{c}}ois}, {Gallou}, {Gharsa}, {Girard}, {Glindemann}, {Guiderdoni},
  {Hahn}, {Hansali}, {Hofmann}, {Jarno}, {Kelz}, {Kiekebusch}, {Knudstrup},
  {Koehler}, {Kollatschny}, {Kosmalski}, {Laurent}, {Le Floch}, {Lilly}, {Lizon
  {\`a} L'Allemand}, {Loupias}, {Manescau}, {Monstein}, {Nicklas}, {Niemeyer},
  {Olaya}, {Palsa}, {Par{\`e}s}, {Pasquini}, {P{\'e}contal-Rousset}, {Pello},
  {Petit}, {Piqueras}, {Popow}, {Reiss}, {Remillieux}, {Renault}, {Rhode},
  {Richard}, {Roth}, {Rupprecht}, {Schaye}, {Slezak}, {Soucail}, {Steinmetz},
  {Streicher}, {Stuik}, {Valentin}, {Vernet}, {Weilbacher}, {Wisotzki},
  {Yerle}, \& {Zins}}]{Bacon_MUSE}
{Bacon}, R., {Accardo}, M., {Adjali}, L., {et~al.} 2012, The Messenger, 147, 4

\bibitem[{{Barrow} {et~al.}(2020){Barrow}, {Robertson}, {Ellis}, {Nakajima},
  {Saxena}, {Stark}, \& {Tang}}]{barrow20}
{Barrow}, K. S.~S., {Robertson}, B.~E., {Ellis}, R.~S., {et~al.} 2020, \apjl,
  902, L39

\bibitem[{{Bastian}(2008)}]{bastian08}
{Bastian}, N. 2008, \mnras, 390, 759

\bibitem[{{Bastian} {et~al.}(2006){Bastian}, {Saglia}, {Goudfrooij},
  {Kissler-Patig}, {Maraston}, {Schweizer}, \& {Zoccali}}]{bastian06}
{Bastian}, N., {Saglia}, R.~P., {Goudfrooij}, P., {et~al.} 2006, \aap, 448, 881

\bibitem[{{Behrens} {et~al.}(2014){Behrens}, {Dijkstra}, \&
  {Niemeyer}}]{behrens14}
{Behrens}, C., {Dijkstra}, M., \& {Niemeyer}, J.~C. 2014, \aap, 563, A77

\bibitem[{{Bergamini} {et~al.}(2021){Bergamini}, {Rosati}, {Vanzella},
  {Caminha}, {Grillo}, {Mercurio}, {Meneghetti}, {Angora}, {Calura}, {Nonino},
  \& {Tozzi}}]{bergamini21}
{Bergamini}, P., {Rosati}, P., {Vanzella}, E., {et~al.} 2021, \aap, 645, A140

\bibitem[{{Bik} {et~al.}(2018){Bik}, {{\"O}stlin}, {Menacho}, {Adamo}, {Hayes},
  {Herenz}, \& {Melinder}}]{bik18}
{Bik}, A., {{\"O}stlin}, G., {Menacho}, V., {et~al.} 2018, \aap, 619, A131

\bibitem[{{Bournaud}(2016)}]{bournaud16}
{Bournaud}, F. 2016, {Bulge Growth Through Disc Instabilities in High-Redshift
  Galaxies}, ed. E.~{Laurikainen}, R.~{Peletier}, \& D.~{Gadotti}, Vol. 418,
  355

\bibitem[{{Bruzual} \& {Charlot}(2003)}]{Bruzual_2003}
{Bruzual}, G. \& {Charlot}, S. 2003, \mnras, 344, 1000

\bibitem[{{Cabrera-Ziri} {et~al.}(2016){Cabrera-Ziri}, {Bastian}, {Hilker},
  {Davies}, {Schweizer}, {Kruijssen}, {Mej{\'\i}a-Narv{\'a}ez}, {Niederhofer},
  {Brandt}, {Rejkuba}, {Bruzual}, \& {Magris}}]{cabrera16}
{Cabrera-Ziri}, I., {Bastian}, N., {Hilker}, M., {et~al.} 2016, \mnras, 457,
  809

\bibitem[{{Cabrera-Ziri} {et~al.}(2015){Cabrera-Ziri}, {Bastian}, {Longmore},
  {Brogan}, {Hollyhead}, {Larsen}, {Whitmore}, {Johnson}, {Chandar}, {Henshaw},
  {Davies}, \& {Hibbard}}]{cabrera15}
{Cabrera-Ziri}, I., {Bastian}, N., {Longmore}, S.~N., {et~al.} 2015, \mnras,
  448, 2224

\bibitem[{{Calzetti} {et~al.}(2000){Calzetti}, {Armus}, {Bohlin}, {Kinney},
  {Koornneef}, \& {Storchi-Bergmann}}]{Calzetti_2000}
{Calzetti}, D., {Armus}, L., {Bohlin}, R.~C., {et~al.} 2000, \apj, 533, 682

\bibitem[{{Calzetti} {et~al.}(2015{\natexlab{a}}){Calzetti}, {Johnson},
  {Adamo}, {Gallagher}, {Andrews}, {Smith}, {Clayton}, {Lee}, {Sabbi}, {Ubeda},
  {Kim}, {Ryon}, {Thilker}, {Bright}, {Zackrisson}, {Kennicutt}, {de Mink},
  {Whitmore}, {Aloisi}, {Chandar}, {Cignoni}, {Cook}, {Dale}, {Elmegreen},
  {Elmegreen}, {Evans}, {Fumagalli}, {Gouliermis}, {Grasha}, {Grebel},
  {Krumholz}, {Walterbos}, {Wofford}, {Brown}, {Christian}, {Dobbs}, {Herrero},
  {Kahre}, {Messa}, {Nair}, {Nota}, {{\"O}stlin}, {Pellerin}, {Sacchi},
  {Schaerer}, \& {Tosi}}]{calzetti15_ngc5253}
{Calzetti}, D., {Johnson}, K.~E., {Adamo}, A., {et~al.} 2015{\natexlab{a}},
  \apj, 811, 75

\bibitem[{{Calzetti} {et~al.}(2015{\natexlab{b}}){Calzetti}, {Lee}, {Sabbi},
  {Adamo}, {Smith}, {Andrews}, {Ubeda}, {Bright}, {Thilker}, {Aloisi}, {Brown},
  {Chandar}, {Christian}, {Cignoni}, {Clayton}, {da Silva}, {de Mink}, {Dobbs},
  {Elmegreen}, {Elmegreen}, {Evans}, {Fumagalli}, {Gallagher}, {Gouliermis},
  {Grebel}, {Herrero}, {Hunter}, {Johnson}, {Kennicutt}, {Kim}, {Krumholz},
  {Lennon}, {Levay}, {Martin}, {Nair}, {Nota}, {{\"O}stlin}, {Pellerin},
  {Prieto}, {Regan}, {Ryon}, {Schaerer}, {Schiminovich}, {Tosi}, {Van Dyk},
  {Walterbos}, {Whitmore}, \& {Wofford}}]{calzetti15}
{Calzetti}, D., {Lee}, J.~C., {Sabbi}, E., {et~al.} 2015{\natexlab{b}}, \aj,
  149, 51

\bibitem[{{Caminha} {et~al.}(2017){Caminha}, {Grillo}, {Rosati}, {Balestra},
  {Mercurio}, {Vanzella}, {Biviano}, {Caputi}, {Delgado-Correal}, {Karman},
  {Lombardi}, {Meneghetti}, {Sartoris}, \& {Tozzi}}]{Caminha_macs0416}
{Caminha}, G.~B., {Grillo}, C., {Rosati}, P., {et~al.} 2017, \aap, 600, A90

\bibitem[{{Carnall} {et~al.}(2019){Carnall}, {Leja}, {Johnson}, {McLure},
  {Dunlop}, \& {Conroy}}]{carnall19}
{Carnall}, A.~C., {Leja}, J., {Johnson}, B.~D., {et~al.} 2019, \apj, 873, 44

\bibitem[{{Castellano} {et~al.}(2016){Castellano}, {Amor{\'\i}n}, {Merlin},
  {Fontana}, {McLure}, {M{\'a}rmol-Queralt{\'o}}, {Mortlock}, {Parsa},
  {Dunlop}, {Elbaz}, {Balestra}, {Boucaud}, {Bourne}, {Boutsia}, {Brammer},
  {Bruce}, {Buitrago}, {Capak}, {Cappelluti}, {Ciesla}, {Comastri}, {Cullen},
  {Derriere}, {Faber}, {Giallongo}, {Grazian}, {Grillo}, {Mercurio},
  {Micha{\l}owski}, {Nonino}, {Paris}, {Pentericci}, {Pilo}, {Rosati},
  {Santini}, {Schreiber}, {Shu}, \& {Wang}}]{castellano16}
{Castellano}, M., {Amor{\'\i}n}, R., {Merlin}, E., {et~al.} 2016, \aap, 590,
  A31

\bibitem[{{Castellano} {et~al.}(2017){Castellano}, {Pentericci}, {Fontana},
  {Vanzella}, {Merlin}, {De Barros}, {Amorin}, {Caputi}, {Cristiani},
  {Finkelstein}, {Giallongo}, {Grazian}, {Koekemoer}, {Maiolino}, {Paris},
  {Pilo}, {Santini}, \& {Yan}}]{castellano17}
{Castellano}, M., {Pentericci}, L., {Fontana}, A., {et~al.} 2017, \apj, 839, 73

\bibitem[{{Cava} {et~al.}(2018){Cava}, {Schaerer}, {Richard},
  {P{\'e}rez-Gonz{\'a}lez}, {Dessauges-Zavadsky}, {Mayer}, \&
  {Tamburello}}]{cava18}
{Cava}, A., {Schaerer}, D., {Richard}, J., {et~al.} 2018, Nature Astronomy, 2,
  76

\bibitem[{{Chevallard} {et~al.}(2018){Chevallard}, {Charlot}, {Senchyna},
  {Stark}, {Vidal-Garc{\'\i}a}, {Feltre}, {Gutkin}, {Jones}, {Mainali}, \&
  {Wofford}}]{chevallard18}
{Chevallard}, J., {Charlot}, S., {Senchyna}, P., {et~al.} 2018, \mnras, 479,
  3264

\bibitem[{{Chisholm} {et~al.}(2019){Chisholm}, {Rigby}, {Bayliss}, {Berg},
  {Dahle}, {Gladders}, \& {Sharon}}]{chisholm19}
{Chisholm}, J., {Rigby}, J.~R., {Bayliss}, M., {et~al.} 2019, \apj, 882, 182

\bibitem[{{Dahle} {et~al.}(2016){Dahle}, {Aghanim}, {Guennou}, {Hudelot},
  {Kneissl}, {Pointecouteau}, {Beelen}, {Bayliss}, {Douspis}, {Nesvadba},
  {Hempel}, {Gronke}, {Burenin}, {Dole}, {Harrison}, {Mazzotta}, \&
  {Sunyaev}}]{dahle16}
{Dahle}, H., {Aghanim}, N., {Guennou}, L., {et~al.} 2016, \aap, 590, L4

\bibitem[{{De Barros} {et~al.}(2019){De Barros}, {Oesch}, {Labb{\'e}},
  {Stefanon}, {Gonz{\'a}lez}, {Smit}, {Bouwens}, \& {Illingworth}}]{barros2019}
{De Barros}, S., {Oesch}, P.~A., {Labb{\'e}}, I., {et~al.} 2019, \mnras, 489,
  2355

\bibitem[{{de Barros} {et~al.}(2016){de Barros}, {Vanzella}, {Amor{\'\i}n},
  {Castellano}, {Siana}, {Grazian}, {Suh}, {Balestra}, {Vignali}, {Verhamme},
  {Zamorani}, {Mignoli}, {Hasinger}, {Comastri}, {Pentericci},
  {P{\'e}rez-Montero}, {Fontana}, {Giavalisco}, \& {Gilli}}]{barros16}
{de Barros}, S., {Vanzella}, E., {Amor{\'\i}n}, R., {et~al.} 2016, \aap, 585,
  A51

\bibitem[{{Dessauges-Zavadsky} {et~al.}(2019){Dessauges-Zavadsky}, {Richard},
  {Combes}, {Schaerer}, {Rujopakarn}, {Mayer}, {Cava}, {Boone}, {Egami},
  {Kneib}, {P{\'e}rez-Gonz{\'a}lez}, {Pfenniger}, {Rawle}, {Teyssier}, \& {van
  der Werf}}]{mirka19}
{Dessauges-Zavadsky}, M., {Richard}, J., {Combes}, F., {et~al.} 2019, Nature
  Astronomy, 3, 1115

\bibitem[{{Dessauges-Zavadsky} {et~al.}(2017){Dessauges-Zavadsky}, {Schaerer},
  {Cava}, {Mayer}, \& {Tamburello}}]{mirka17}
{Dessauges-Zavadsky}, M., {Schaerer}, D., {Cava}, A., {Mayer}, L., \&
  {Tamburello}, V. 2017, \apjl, 836, L22

\bibitem[{{Du} {et~al.}(2020){Du}, {Shapley}, {Tang}, {Stark}, {Martin},
  {Mobasher}, {Topping}, \& {Chevallard}}]{du20_z2_analogs_toEOR}
{Du}, X., {Shapley}, A.~E., {Tang}, M., {et~al.} 2020, \apj, 890, 65

\bibitem[{{Elmegreen} {et~al.}(2007){Elmegreen}, {Elmegreen}, {Ravindranath},
  \& {Coe}}]{elmegreen07}
{Elmegreen}, D.~M., {Elmegreen}, B.~G., {Ravindranath}, S., \& {Coe}, D.~A.
  2007, \apj, 658, 763

\bibitem[{{Elmegreen} {et~al.}(2020){Elmegreen}, {Elmegreen}, {Whitmore},
  {Chandar}, {Calzetti}, {Lee}, {White}, {Cook}, {Ubeda}, {Mok}, \&
  {Linden}}]{elmegreen20}
{Elmegreen}, D.~M., {Elmegreen}, B.~G., {Whitmore}, B.~C., {et~al.} 2020, arXiv
  e-prints, arXiv:2012.10765

\bibitem[{{Endsley} {et~al.}(2021){Endsley}, {Stark}, {Chevallard}, \&
  {Charlot}}]{endsley21}
{Endsley}, R., {Stark}, D.~P., {Chevallard}, J., \& {Charlot}, S. 2021, \mnras,
  500, 5229

\bibitem[{{Faure} {et~al.}(2021){Faure}, {Bournaud}, {Fensch}, {Daddi},
  {Behrendt}, {Burkert}, \& {Richard}}]{faure21}
{Faure}, B., {Bournaud}, F., {Fensch}, J., {et~al.} 2021, arXiv e-prints,
  arXiv:2101.11013

\bibitem[{{Finkelstein} {et~al.}(2013){Finkelstein}, {Papovich}, {Dickinson},
  {Song}, {Tilvi}, {Koekemoer}, {Finkelstein}, {Mobasher}, {Ferguson},
  {Giavalisco}, {Reddy}, {Ashby}, {Dekel}, {Fazio}, {Fontana}, {Grogin},
  {Huang}, {Kocevski}, {Rafelski}, {Weiner}, \&
  {Willner}}]{finkelstein13nature}
{Finkelstein}, S.~L., {Papovich}, C., {Dickinson}, M., {et~al.} 2013, \nat,
  502, 524

\bibitem[{{Fontana} {et~al.}(2000){Fontana}, {D'Odorico}, {Poli}, {Giallongo},
  {Arnouts}, {Cristiani}, {Moorwood}, \& {Saracco}}]{fontana2000}
{Fontana}, A., {D'Odorico}, S., {Poli}, F., {et~al.} 2000, \aj, 120, 2206

\bibitem[{{F{\"o}rster Schreiber} {et~al.}(2011){F{\"o}rster Schreiber},
  {Shapley}, {Genzel}, {Bouch{\'e}}, {Cresci}, {Davies}, {Erb}, {Genel},
  {Lutz}, {Newman}, {Shapiro}, {Steidel}, {Sternberg}, \& {Tacconi}}]{forser11}
{F{\"o}rster Schreiber}, N.~M., {Shapley}, A.~E., {Genzel}, R., {et~al.} 2011,
  \apj, 739, 45

\bibitem[{{Gaia Collaboration} {et~al.}(2018){Gaia Collaboration}, {Brown},
  {Vallenari}, {Prusti}, {de Bruijne}, {Babusiaux}, {Bailer-Jones}, {Biermann},
  {Evans}, {Eyer}, {Jansen}, {Jordi}, {Klioner}, {Lammers}, {Lindegren},
  {Luri}, {Mignard}, {Panem}, {Pourbaix}, {Randich}, {Sartoretti}, {Siddiqui},
  {Soubiran}, {van Leeuwen}, {Walton}, {Arenou}, {Bastian}, {Cropper},
  {Drimmel}, {Katz}, {Lattanzi}, {Bakker}, {Cacciari}, {Casta{\~n}eda},
  {Chaoul}, {Cheek}, {De Angeli}, {Fabricius}, {Guerra}, {Holl}, {Masana},
  {Messineo}, {Mowlavi}, {Nienartowicz}, {Panuzzo}, {Portell}, {Riello},
  {Seabroke}, {Tanga}, {Th{\'e}venin}, {Gracia-Abril}, {Comoretto},
  {Garcia-Reinaldos}, {Teyssier}, {Altmann}, {Andrae}, {Audard},
  {Bellas-Velidis}, {Benson}, {Berthier}, {Blomme}, {Burgess}, {Busso},
  {Carry}, {Cellino}, {Clementini}, {Clotet}, {Creevey}, {Davidson}, {De
  Ridder}, {Delchambre}, {Dell'Oro}, {Ducourant},
  {Fern{\'a}ndez-Hern{\'a}ndez}, {Fouesneau}, {Fr{\'e}mat}, {Galluccio},
  {Garc{\'\i}a-Torres}, {Gonz{\'a}lez-N{\'u}{\~n}ez}, {Gonz{\'a}lez-Vidal},
  {Gosset}, {Guy}, {Halbwachs}, {Hambly}, {Harrison}, {Hern{\'a}ndez},
  {Hestroffer}, {Hodgkin}, {Hutton}, {Jasniewicz}, {Jean-Antoine-Piccolo},
  {Jordan}, {Korn}, {Krone-Martins}, {Lanzafame}, {Lebzelter}, {L{\"o}ffler},
  {Manteiga}, {Marrese}, {Mart{\'\i}n-Fleitas}, {Moitinho}, {Mora}, {Muinonen},
  {Osinde}, {Pancino}, {Pauwels}, {Petit}, {Recio-Blanco}, {Richards},
  {Rimoldini}, {Robin}, {Sarro}, {Siopis}, {Smith}, {Sozzetti}, {S{\"u}veges},
  {Torra}, {van Reeven}, {Abbas}, {Abreu Aramburu}, {Accart}, {Aerts},
  {Altavilla}, {{\'A}lvarez}, {Alvarez}, {Alves}, {Anderson}, {Andrei},
  {Anglada Varela}, {Antiche}, {Antoja}, {Arcay}, {Astraatmadja}, {Bach},
  {Baker}, {Balaguer-N{\'u}{\~n}ez}, {Balm}, {Barache}, {Barata}, {Barbato},
  {Barblan}, {Barklem}, {Barrado}, {Barros}, {Barstow}, {Bartholom{\'e}
  Mu{\~n}oz}, {Bassilana}, {Becciani}, {Bellazzini}, {Berihuete}, {Bertone},
  {Bianchi}, {Bienaym{\'e}}, {Blanco-Cuaresma}, {Boch}, {Boeche}, {Bombrun},
  {Borrachero}, {Bossini}, {Bouquillon}, {Bourda}, {Bragaglia}, {Bramante},
  {Breddels}, {Bressan}, {Brouillet}, {Br{\"u}semeister}, {Brugaletta},
  {Bucciarelli}, {Burlacu}, {Busonero}, {Butkevich}, {Buzzi}, {Caffau},
  {Cancelliere}, {Cannizzaro}, {Cantat-Gaudin}, {Carballo}, {Carlucci},
  {Carrasco}, {Casamiquela}, {Castellani}, {Castro-Ginard}, {Charlot},
  {Chemin}, {Chiavassa}, {Cocozza}, {Costigan}, {Cowell}, {Crifo}, {Crosta},
  {Crowley}, {Cuypers}, {Dafonte}, {Damerdji}, {Dapergolas}, {David}, {David},
  {de Laverny}, {De Luise}, {De March}, {de Martino}, {de Souza}, {de Torres},
  {Debosscher}, {del Pozo}, {Delbo}, {Delgado}, {Delgado}, {Di Matteo},
  {Diakite}, {Diener}, {Distefano}, {Dolding}, {Drazinos}, {Dur{\'a}n},
  {Edvardsson}, {Enke}, {Eriksson}, {Esquej}, {Eynard Bontemps}, {Fabre},
  {Fabrizio}, {Faigler}, {Falc{\~a}o}, {Farr{\`a}s Casas}, {Federici},
  {Fedorets}, {Fernique}, {Figueras}, {Filippi}, {Findeisen}, {Fonti},
  {Fraile}, {Fraser}, {Fr{\'e}zouls}, {Gai}, {Galleti}, {Garabato},
  {Garc{\'\i}a-Sedano}, {Garofalo}, {Garralda}, {Gavel}, {Gavras}, {Gerssen},
  {Geyer}, {Giacobbe}, {Gilmore}, {Girona}, {Giuffrida}, {Glass}, {Gomes},
  {Granvik}, {Gueguen}, {Guerrier}, {Guiraud}, {Guti{\'e}rrez-S{\'a}nchez},
  {Haigron}, {Hatzidimitriou}, {Hauser}, {Haywood}, {Heiter}, {Helmi}, {Heu},
  {Hilger}, {Hobbs}, {Hofmann}, {Holland}, {Huckle}, {Hypki}, {Icardi},
  {Jan{\ss}en}, {Jevardat de Fombelle}, {Jonker}, {Juh{\'a}sz}, {Julbe},
  {Karampelas}, {Kewley}, {Klar}, {Kochoska}, {Kohley}, {Kolenberg},
  {Kontizas}, {Kontizas}, {Koposov}, {Kordopatis}, {Kostrzewa-Rutkowska},
  {Koubsky}, {Lambert}, {Lanza}, {Lasne}, {Lavigne}, {Le Fustec}, {Le
  Poncin-Lafitte}, {Lebreton}, {Leccia}, {Leclerc}, {Lecoeur-Taibi},
  {Lenhardt}, {Leroux}, {Liao}, {Licata}, {Lindstr{\o}m}, {Lister}, {Livanou},
  {Lobel}, {L{\'o}pez}, {Managau}, {Mann}, {Mantelet}, {Marchal}, {Marchant},
  {Marconi}, {Marinoni}, {Marschalk{\'o}}, {Marshall}, {Martino}, {Marton},
  {Mary}, {Massari}, {Matijevi{\v{c}}}, {Mazeh}, {McMillan}, {Messina},
  {Michalik}, {Millar}, {Molina}, {Molinaro}, {Moln{\'a}r}, {Montegriffo},
  {Mor}, {Morbidelli}, {Morel}, {Morris}, {Mulone}, {Muraveva}, {Musella},
  {Nelemans}, {Nicastro}, {Noval}, {O'Mullane}, {Ord{\'e}novic},
  {Ord{\'o}{\~n}ez-Blanco}, {Osborne}, {Pagani}, {Pagano}, {Pailler},
  {Palacin}, {Palaversa}, {Panahi}, {Pawlak}, {Piersimoni}, {Pineau}, {Plachy},
  {Plum}, {Poggio}, {Poujoulet}, {Pr{\v{s}}a}, {Pulone}, {Racero}, {Ragaini},
  {Rambaux}, {Ramos-Lerate}, {Regibo}, {Reyl{\'e}}, {Riclet}, {Ripepi}, {Riva},
  {Rivard}, {Rixon}, {Roegiers}, {Roelens}, {Romero-G{\'o}mez}, {Rowell},
  {Royer}, {Ruiz-Dern}, {Sadowski}, {Sagrist{\`a} Sell{\'e}s}, {Sahlmann},
  {Salgado}, {Salguero}, {Sanna}, {Santana-Ros}, {Sarasso}, {Savietto},
  {Schultheis}, {Sciacca}, {Segol}, {Segovia}, {S{\'e}gransan}, {Shih},
  {Siltala}, {Silva}, {Smart}, {Smith}, {Solano}, {Solitro}, {Sordo}, {Soria
  Nieto}, {Souchay}, {Spagna}, {Spoto}, {Stampa}, {Steele},
  {Steidelm{\"u}ller}, {Stephenson}, {Stoev}, {Suess}, {Surdej}, {Szabados},
  {Szegedi-Elek}, {Tapiador}, {Taris}, {Tauran}, {Taylor}, {Teixeira},
  {Terrett}, {Teyssandier}, {Thuillot}, {Titarenko}, {Torra Clotet}, {Turon},
  {Ulla}, {Utrilla}, {Uzzi}, {Vaillant}, {Valentini}, {Valette}, {van Elteren},
  {Van Hemelryck}, {van Leeuwen}, {Vaschetto}, {Vecchiato}, {Veljanoski},
  {Viala}, {Vicente}, {Vogt}, {von Essen}, {Voss}, {Votruba}, {Voutsinas},
  {Walmsley}, {Weiler}, {Wertz}, {Wevers}, {Wyrzykowski}, {Yoldas},
  {{\v{Z}}erjal}, {Ziaeepour}, {Zorec}, {Zschocke}, {Zucker}, {Zurbach}, \&
  {Zwitter}}]{gaia2018}
{Gaia Collaboration}, {Brown}, A.~G.~A., {Vallenari}, A., {et~al.} 2018, \aap,
  616, A1

\bibitem[{{Gaia Collaboration} {et~al.}(2016){Gaia Collaboration}, {Prusti},
  {de Bruijne}, {Brown}, {Vallenari}, {Babusiaux}, {Bailer-Jones}, {Bastian},
  {Biermann}, {Evans}, {Eyer}, {Jansen}, {Jordi}, {Klioner}, {Lammers},
  {Lindegren}, {Luri}, {Mignard}, {Milligan}, {Panem}, {Poinsignon},
  {Pourbaix}, {Randich}, {Sarri}, {Sartoretti}, {Siddiqui}, {Soubiran},
  {Valette}, {van Leeuwen}, {Walton}, {Aerts}, {Arenou}, {Cropper}, {Drimmel},
  {H{\o}g}, {Katz}, {Lattanzi}, {O'Mullane}, {Grebel}, {Holland}, {Huc},
  {Passot}, {Bramante}, {Cacciari}, {Casta{\~n}eda}, {Chaoul}, {Cheek}, {De
  Angeli}, {Fabricius}, {Guerra}, {Hern{\'a}ndez}, {Jean-Antoine-Piccolo},
  {Masana}, {Messineo}, {Mowlavi}, {Nienartowicz}, {Ord{\'o}{\~n}ez-Blanco},
  {Panuzzo}, {Portell}, {Richards}, {Riello}, {Seabroke}, {Tanga},
  {Th{\'e}venin}, {Torra}, {Els}, {Gracia-Abril}, {Comoretto},
  {Garcia-Reinaldos}, {Lock}, {Mercier}, {Altmann}, {Andrae}, {Astraatmadja},
  {Bellas-Velidis}, {Benson}, {Berthier}, {Blomme}, {Busso}, {Carry},
  {Cellino}, {Clementini}, {Cowell}, {Creevey}, {Cuypers}, {Davidson}, {De
  Ridder}, {de Torres}, {Delchambre}, {Dell'Oro}, {Ducourant}, {Fr{\'e}mat},
  {Garc{\'\i}a-Torres}, {Gosset}, {Halbwachs}, {Hambly}, {Harrison}, {Hauser},
  {Hestroffer}, {Hodgkin}, {Huckle}, {Hutton}, {Jasniewicz}, {Jordan},
  {Kontizas}, {Korn}, {Lanzafame}, {Manteiga}, {Moitinho}, {Muinonen},
  {Osinde}, {Pancino}, {Pauwels}, {Petit}, {Recio-Blanco}, {Robin}, {Sarro},
  {Siopis}, {Smith}, {Smith}, {Sozzetti}, {Thuillot}, {van Reeven}, {Viala},
  {Abbas}, {Abreu Aramburu}, {Accart}, {Aguado}, {Allan}, {Allasia},
  {Altavilla}, {{\'A}lvarez}, {Alves}, {Anderson}, {Andrei}, {Anglada Varela},
  {Antiche}, {Antoja}, {Ant{\'o}n}, {Arcay}, {Atzei}, {Ayache}, {Bach},
  {Baker}, {Balaguer-N{\'u}{\~n}ez}, {Barache}, {Barata}, {Barbier}, {Barblan},
  {Baroni}, {Barrado y Navascu{\'e}s}, {Barros}, {Barstow}, {Becciani},
  {Bellazzini}, {Bellei}, {Bello Garc{\'\i}a}, {Belokurov}, {Bendjoya},
  {Berihuete}, {Bianchi}, {Bienaym{\'e}}, {Billebaud}, {Blagorodnova},
  {Blanco-Cuaresma}, {Boch}, {Bombrun}, {Borrachero}, {Bouquillon}, {Bourda},
  {Bouy}, {Bragaglia}, {Breddels}, {Brouillet}, {Br{\"u}semeister},
  {Bucciarelli}, {Budnik}, {Burgess}, {Burgon}, {Burlacu}, {Busonero}, {Buzzi},
  {Caffau}, {Cambras}, {Campbell}, {Cancelliere}, {Cantat-Gaudin}, {Carlucci},
  {Carrasco}, {Castellani}, {Charlot}, {Charnas}, {Charvet}, {Chassat},
  {Chiavassa}, {Clotet}, {Cocozza}, {Collins}, {Collins}, {Costigan}, {Crifo},
  {Cross}, {Crosta}, {Crowley}, {Dafonte}, {Damerdji}, {Dapergolas}, {David},
  {David}, {De Cat}, {de Felice}, {de Laverny}, {De Luise}, {De March}, {de
  Martino}, {de Souza}, {Debosscher}, {del Pozo}, {Delbo}, {Delgado},
  {Delgado}, {di Marco}, {Di Matteo}, {Diakite}, {Distefano}, {Dolding}, {Dos
  Anjos}, {Drazinos}, {Dur{\'a}n}, {Dzigan}, {Ecale}, {Edvardsson}, {Enke},
  {Erdmann}, {Escolar}, {Espina}, {Evans}, {Eynard Bontemps}, {Fabre},
  {Fabrizio}, {Faigler}, {Falc{\~a}o}, {Farr{\`a}s Casas}, {Faye}, {Federici},
  {Fedorets}, {Fern{\'a}ndez-Hern{\'a}ndez}, {Fernique}, {Fienga}, {Figueras},
  {Filippi}, {Findeisen}, {Fonti}, {Fouesneau}, {Fraile}, {Fraser}, {Fuchs},
  {Furnell}, {Gai}, {Galleti}, {Galluccio}, {Garabato}, {Garc{\'\i}a-Sedano},
  {Gar{\'e}}, {Garofalo}, {Garralda}, {Gavras}, {Gerssen}, {Geyer}, {Gilmore},
  {Girona}, {Giuffrida}, {Gomes}, {Gonz{\'a}lez-Marcos},
  {Gonz{\'a}lez-N{\'u}{\~n}ez}, {Gonz{\'a}lez-Vidal}, {Granvik}, {Guerrier},
  {Guillout}, {Guiraud}, {G{\'u}rpide}, {Guti{\'e}rrez-S{\'a}nchez}, {Guy},
  {Haigron}, {Hatzidimitriou}, {Haywood}, {Heiter}, {Helmi}, {Hobbs},
  {Hofmann}, {Holl}, {Holland}, {Hunt}, {Hypki}, {Icardi}, {Irwin}, {Jevardat
  de Fombelle}, {Jofr{\'e}}, {Jonker}, {Jorissen}, {Julbe}, {Karampelas},
  {Kochoska}, {Kohley}, {Kolenberg}, {Kontizas}, {Koposov}, {Kordopatis},
  {Koubsky}, {Kowalczyk}, {Krone-Martins}, {Kudryashova}, {Kull}, {Bachchan},
  {Lacoste-Seris}, {Lanza}, {Lavigne}, {Le Poncin-Lafitte}, {Lebreton},
  {Lebzelter}, {Leccia}, {Leclerc}, {Lecoeur-Taibi}, {Lemaitre}, {Lenhardt},
  {Leroux}, {Liao}, {Licata}, {Lindstr{\o}m}, {Lister}, {Livanou}, {Lobel},
  {L{\"o}ffler}, {L{\'o}pez}, {Lopez-Lozano}, {Lorenz}, {Loureiro},
  {MacDonald}, {Magalh{\~a}es Fernandes}, {Managau}, {Mann}, {Mantelet},
  {Marchal}, {Marchant}, {Marconi}, {Marie}, {Marinoni}, {Marrese},
  {Marschalk{\'o}}, {Marshall}, {Mart{\'\i}n-Fleitas}, {Martino}, {Mary},
  {Matijevi{\v{c}}}, {Mazeh}, {McMillan}, {Messina}, {Mestre}, {Michalik},
  {Millar}, {Miranda}, {Molina}, {Molinaro}, {Molinaro}, {Moln{\'a}r},
  {Moniez}, {Montegriffo}, {Monteiro}, {Mor}, {Mora}, {Morbidelli}, {Morel},
  {Morgenthaler}, {Morley}, {Morris}, {Mulone}, {Muraveva}, {Musella},
  {Narbonne}, {Nelemans}, {Nicastro}, {Noval}, {Ord{\'e}novic},
  {Ordieres-Mer{\'e}}, {Osborne}, {Pagani}, {Pagano}, {Pailler}, {Palacin},
  {Palaversa}, {Parsons}, {Paulsen}, {Pecoraro}, {Pedrosa}, {Pentik{\"a}inen},
  {Pereira}, {Pichon}, {Piersimoni}, {Pineau}, {Plachy}, {Plum}, {Poujoulet},
  {Pr{\v{s}}a}, {Pulone}, {Ragaini}, {Rago}, {Rambaux}, {Ramos-Lerate},
  {Ranalli}, {Rauw}, {Read}, {Regibo}, {Renk}, {Reyl{\'e}}, {Ribeiro},
  {Rimoldini}, {Ripepi}, {Riva}, {Rixon}, {Roelens}, {Romero-G{\'o}mez},
  {Rowell}, {Royer}, {Rudolph}, {Ruiz-Dern}, {Sadowski}, {Sagrist{\`a}
  Sell{\'e}s}, {Sahlmann}, {Salgado}, {Salguero}, {Sarasso}, {Savietto},
  {Schnorhk}, {Schultheis}, {Sciacca}, {Segol}, {Segovia}, {Segransan},
  {Serpell}, {Shih}, {Smareglia}, {Smart}, {Smith}, {Solano}, {Solitro},
  {Sordo}, {Soria Nieto}, {Souchay}, {Spagna}, {Spoto}, {Stampa}, {Steele},
  {Steidelm{\"u}ller}, {Stephenson}, {Stoev}, {Suess}, {S{\"u}veges}, {Surdej},
  {Szabados}, {Szegedi-Elek}, {Tapiador}, {Taris}, {Tauran}, {Taylor},
  {Teixeira}, {Terrett}, {Tingley}, {Trager}, {Turon}, {Ulla}, {Utrilla},
  {Valentini}, {van Elteren}, {Van Hemelryck}, {van Leeuwen}, {Varadi},
  {Vecchiato}, {Veljanoski}, {Via}, {Vicente}, {Vogt}, {Voss}, {Votruba},
  {Voutsinas}, {Walmsley}, {Weiler}, {Weingrill}, {Werner}, {Wevers},
  {Whitehead}, {Wyrzykowski}, {Yoldas}, {{\v{Z}}erjal}, {Zucker}, {Zurbach},
  {Zwitter}, {Alecu}, {Allen}, {Allende Prieto}, {Amorim},
  {Anglada-Escud{\'e}}, {Arsenijevic}, {Azaz}, {Balm}, {Beck}, {Bernstein},
  {Bigot}, {Bijaoui}, {Blasco}, {Bonfigli}, {Bono}, {Boudreault}, {Bressan},
  {Brown}, {Brunet}, {Bunclark}, {Buonanno}, {Butkevich}, {Carret}, {Carrion},
  {Chemin}, {Ch{\'e}reau}, {Corcione}, {Darmigny}, {de Boer}, {de Teodoro}, {de
  Zeeuw}, {Delle Luche}, {Domingues}, {Dubath}, {Fodor}, {Fr{\'e}zouls},
  {Fries}, {Fustes}, {Fyfe}, {Gallardo}, {Gallegos}, {Gardiol}, {Gebran},
  {Gomboc}, {G{\'o}mez}, {Grux}, {Gueguen}, {Heyrovsky}, {Hoar}, {Iannicola},
  {Isasi Parache}, {Janotto}, {Joliet}, {Jonckheere}, {Keil}, {Kim},
  {Klagyivik}, {Klar}, {Knude}, {Kochukhov}, {Kolka}, {Kos}, {Kutka}, {Lainey},
  {LeBouquin}, {Liu}, {Loreggia}, {Makarov}, {Marseille}, {Martayan},
  {Martinez-Rubi}, {Massart}, {Meynadier}, {Mignot}, {Munari}, {Nguyen},
  {Nordlander}, {Ocvirk}, {O'Flaherty}, {Olias Sanz}, {Ortiz}, {Osorio},
  {Oszkiewicz}, {Ouzounis}, {Palmer}, {Park}, {Pasquato}, {Peltzer}, {Peralta},
  {P{\'e}turaud}, {Pieniluoma}, {Pigozzi}, {Poels}, {Prat}, {Prod'homme},
  {Raison}, {Rebordao}, {Risquez}, {Rocca-Volmerange}, {Rosen}, {Ruiz-Fuertes},
  {Russo}, {Sembay}, {Serraller Vizcaino}, {Short}, {Siebert}, {Silva},
  {Sinachopoulos}, {Slezak}, {Soffel}, {Sosnowska}, {Strai{\v{z}}ys}, {ter
  Linden}, {Terrell}, {Theil}, {Tiede}, {Troisi}, {Tsalmantza}, {Tur},
  {Vaccari}, {Vachier}, {Valles}, {Van Hamme}, {Veltz}, {Virtanen}, {Wallut},
  {Wichmann}, {Wilkinson}, {Ziaeepour}, \& {Zschocke}}]{gaia2016}
{Gaia Collaboration}, {Prusti}, T., {de Bruijne}, J.~H.~J., {et~al.} 2016,
  \aap, 595, A1

\bibitem[{{Georgiev} {et~al.}(2009){Georgiev}, {Hilker}, {Puzia}, {Goudfrooij},
  \& {Baumgardt}}]{georgiev09}
{Georgiev}, I.~Y., {Hilker}, M., {Puzia}, T.~H., {Goudfrooij}, P., \&
  {Baumgardt}, H. 2009, \mnras, 396, 1075

\bibitem[{{Gieles} \& {Portegies Zwart}(2011)}]{gieles11}
{Gieles}, M. \& {Portegies Zwart}, S.~F. 2011, \mnras, 410, L6

\bibitem[{{Gonzaga} {et~al.}(2012){Gonzaga}, {Hack}, {Fruchter}, \&
  {Mack}}]{gonzaga12}
{Gonzaga}, S., {Hack}, W., {Fruchter}, A., \& {Mack}, J. 2012, {The DrizzlePac
  Handbook}

\bibitem[{{Guo} {et~al.}(2018){Guo}, {Rafelski}, {Bell}, {Conselice}, {Dekel},
  {Faber}, {Giavalisco}, {Koekemoer}, {Koo}, {Lu}, {Mandelker}, {Primack},
  {Ceverino}, {de Mello}, {Ferguson}, {Hathi}, {Kocevski}, {Lucas},
  {P{\'e}rez-Gonz{\'a}lez}, {Ravindranath}, {Soto}, {Straughn}, \&
  {Wang}}]{guo18}
{Guo}, Y., {Rafelski}, M., {Bell}, E.~F., {et~al.} 2018, \apj, 853, 108

\bibitem[{{Heckman} {et~al.}(2011){Heckman}, {Borthakur}, {Overzier},
  {Kauffmann}, {Basu-Zych}, {Leitherer}, {Sembach}, {Martin}, {Rich},
  {Schiminovich}, \& {Seibert}}]{heckman11}
{Heckman}, T.~M., {Borthakur}, S., {Overzier}, R., {et~al.} 2011, \apj, 730, 5

\bibitem[{{Hunter}(2007)}]{matplotlib2007}
{Hunter}, J.~D. 2007, Computing in Science and Engineering, 9, 90

\bibitem[{{Izotov} {et~al.}(2021{\natexlab{a}}){Izotov}, {Guseva}, {Fricke},
  {Henkel}, {Schaerer}, \& {Thuan}}]{izotov21loz_analogs}
{Izotov}, Y.~I., {Guseva}, N.~G., {Fricke}, K.~J., {et~al.} 2021{\natexlab{a}},
  \aap, 646, A138

\bibitem[{{Izotov} {et~al.}(2016){Izotov}, {Orlitov{\'a}}, {Schaerer}, {Thuan},
  {Verhamme}, {Guseva}, \& {Worseck}}]{izotov16Nature}
{Izotov}, Y.~I., {Orlitov{\'a}}, I., {Schaerer}, D., {et~al.} 2016, \nat, 529,
  178

\bibitem[{{Izotov} {et~al.}(2021{\natexlab{b}}){Izotov}, {Worseck}, {Schaerer},
  {Guseva}, {Chisholm}, {Thuan}, {Fricke}, \& {Verhamme}}]{izotov21lowmass}
{Izotov}, Y.~I., {Worseck}, G., {Schaerer}, D., {et~al.} 2021{\natexlab{b}},
  \mnras, 503, 1734

\bibitem[{{Izotov} {et~al.}(2018){Izotov}, {Worseck}, {Schaerer}, {Guseva},
  {Thuan}, {Fricke}, \& {Orlitov{\'a}}}]{izotov18_O32}
{Izotov}, Y.~I., {Worseck}, G., {Schaerer}, D., {et~al.} 2018, \mnras, 478,
  4851

\bibitem[{{James} {et~al.}(2016){James}, {Auger}, {Aloisi}, {Calzetti}, \&
  {Kewley}}]{james16}
{James}, B.~L., {Auger}, M., {Aloisi}, A., {Calzetti}, D., \& {Kewley}, L.
  2016, \apj, 816, 40

\bibitem[{{Jaskot} {et~al.}(2019){Jaskot}, {Dowd}, {Oey}, {Scarlata}, \&
  {McKinney}}]{jaskot19}
{Jaskot}, A.~E., {Dowd}, T., {Oey}, M.~S., {Scarlata}, C., \& {McKinney}, J.
  2019, \apj, 885, 96

\bibitem[{{Johnson} {et~al.}(2017){Johnson}, {Rigby}, {Sharon}, {Gladders},
  {Florian}, {Bayliss}, {Wuyts}, {Whitaker}, {Livermore}, \&
  {Murray}}]{johnson17}
{Johnson}, T.~L., {Rigby}, J.~R., {Sharon}, K., {et~al.} 2017, \apjl, 843, L21

\bibitem[{{Katz} {et~al.}(2020){Katz}, {{\v{D}}urov{\v{c}}{\'\i}kov{\'a}},
  {Kimm}, {Rosdahl}, {Blaizot}, {Haehnelt}, {Devriendt}, {Slyz}, {Ellis}, \&
  {Laporte}}]{katz20}
{Katz}, H., {{\v{D}}urov{\v{c}}{\'\i}kov{\'a}}, D., {Kimm}, T., {et~al.} 2020,
  \mnras, 498, 164

\bibitem[{{Keenan} {et~al.}(2017){Keenan}, {Oey}, {Jaskot}, \&
  {James}}]{keenan17haro11}
{Keenan}, R.~P., {Oey}, M.~S., {Jaskot}, A.~E., \& {James}, B.~L. 2017, \apj,
  848, 12

\bibitem[{{Krist} {et~al.}(2011){Krist}, {Hook}, \& {Stoehr}}]{krist_2011}
{Krist}, J.~E., {Hook}, R.~N., \& {Stoehr}, F. 2011, in Society of
  Photo-Optical Instrumentation Engineers (SPIE) Conference Series, Vol. 8127,
  Optical Modeling and Performance Predictions V, ed. M.~A. {Kahan}, 81270J

\bibitem[{{Kruijssen}(2012)}]{kruijssen12}
{Kruijssen}, J.~M.~D. 2012, \mnras, 426, 3008

\bibitem[{{Laporte} {et~al.}(2014){Laporte}, {Streblyanska}, {Clement},
  {P{\'e}rez-Fournon}, {Schaerer}, {Atek}, {Boone}, {Kneib}, {Egami},
  {Mart{\'\i}nez-Navajas}, {Marques-Chaves}, {Pell{\'o}}, \&
  {Richard}}]{laporte14}
{Laporte}, N., {Streblyanska}, A., {Clement}, B., {et~al.} 2014, \aap, 562, L8

\bibitem[{{Leitherer} {et~al.}(2014){Leitherer}, {Ekstr{\"o}m}, {Meynet},
  {Schaerer}, {Agienko}, \& {Levesque}}]{leitherer14}
{Leitherer}, C., {Ekstr{\"o}m}, S., {Meynet}, G., {et~al.} 2014, \apjs, 212, 14

\bibitem[{{Livermore} {et~al.}(2015){Livermore}, {Jones}, {Richard}, {Bower},
  {Swinbank}, {Yuan}, {Edge}, {Ellis}, {Kewley}, {Smail}, {Coppin}, \&
  {Ebeling}}]{livermore15}
{Livermore}, R.~C., {Jones}, T.~A., {Richard}, J., {et~al.} 2015, \mnras, 450,
  1812

\bibitem[{{Marques-Chaves} {et~al.}(2021){Marques-Chaves}, {Schaerer},
  {{\'A}lvarez-M{\'a}rquez}, {Colina}, {Dessauges-Zavadsky},
  {P{\'e}rez-Fournon}, {Saldana-Lopez}, \& {Verhamme}}]{RUI2021}
{Marques-Chaves}, R., {Schaerer}, D., {{\'A}lvarez-M{\'a}rquez}, J., {et~al.}
  2021, \mnras, 507, 524

\bibitem[{{Matthee} {et~al.}(2021){Matthee}, {Sobral}, {Hayes}, {Pezzulli},
  {Gronke}, {Schaerer}, {Naidu}, {R{\"o}ttgering}, {Calhau}, {Paulino-Afonso},
  {Santos}, \& {Amor{\'\i}n}}]{matthee21}
{Matthee}, J., {Sobral}, D., {Hayes}, M., {et~al.} 2021, \mnras
  [\eprint[arXiv]{2102.07779}]

\bibitem[{{McLaughlin} \& {van der Marel}(2005)}]{McLaughlin05}
{McLaughlin}, D.~E. \& {van der Marel}, R.~P. 2005, \apjs, 161, 304

\bibitem[{{Meneghetti} {et~al.}(2008){Meneghetti}, {Melchior}, {Grazian}, {De
  Lucia}, {Dolag}, {Bartelmann}, {Heymans}, {Moscardini}, \&
  {Radovich}}]{Meneghetti_2008}
{Meneghetti}, M., {Melchior}, P., {Grazian}, A., {et~al.} 2008, \aap, 482, 403

\bibitem[{{Meneghetti} {et~al.}(2017){Meneghetti}, {Natarajan}, {Coe},
  {Contini}, {De Lucia}, {Giocoli}, {Acebron}, {Borgani}, {Bradac}, {Diego},
  {Hoag}, {Ishigaki}, {Johnson}, {Jullo}, {Kawamata}, {Lam}, {Limousin},
  {Liesenborgs}, {Oguri}, {Sebesta}, {Sharon}, {Williams}, \&
  {Zitrin}}]{Meneghetti_2017}
{Meneghetti}, M., {Natarajan}, P., {Coe}, D., {et~al.} 2017, \mnras, 472, 3177

\bibitem[{{Meneghetti} {et~al.}(2010){Meneghetti}, {Rasia}, {Merten},
  {Bellagamba}, {Ettori}, {Mazzotta}, {Dolag}, \& {Marri}}]{Meneghetti_2010}
{Meneghetti}, M., {Rasia}, E., {Merten}, J., {et~al.} 2010, \aap, 514, A93

\bibitem[{{Merlin} {et~al.}(2019){Merlin}, {Pilo}, {Fontana}, {Castellano},
  {Paris}, {Roscani}, {Santini}, \& {Torelli}}]{merlin19}
{Merlin}, E., {Pilo}, S., {Fontana}, A., {et~al.} 2019, \aap, 622, A169

\bibitem[{{Micheva} {et~al.}(2017){Micheva}, {Oey}, {Jaskot}, \&
  {James}}]{micheva17}
{Micheva}, G., {Oey}, M.~S., {Jaskot}, A.~E., \& {James}, B.~L. 2017, \apj,
  845, 165

\bibitem[{{Naidu} {et~al.}(2020){Naidu}, {Tacchella}, {Mason}, {Bose}, {Oesch},
  \& {Conroy}}]{naidu20}
{Naidu}, R.~P., {Tacchella}, S., {Mason}, C.~A., {et~al.} 2020, \apj, 892, 109

\bibitem[{{{\"O}stlin} {et~al.}(2007){{\"O}stlin}, {Cumming}, \&
  {Bergvall}}]{ostlin07}
{{\"O}stlin}, G., {Cumming}, R.~J., \& {Bergvall}, N. 2007, \aap, 461, 471

\bibitem[{{{\"O}stlin} {et~al.}(2021){{\"O}stlin}, {Rivera-Thorsen}, {Menacho},
  {Hayes}, {Runnholm}, {Micheva}, {Oey}, {Adamo}, {Bik}, {Cannon}, {Gronke},
  {Kunth}, {Laursen}, {Mas-Hesse}, {Melinde}, {Messa}, {Sirressi}, \&
  {Smith}}]{ostlin2haro11_LyC}
{{\"O}stlin}, G., {Rivera-Thorsen}, T.~E., {Menacho}, V., {et~al.} 2021, arXiv
  e-prints, arXiv:2103.15854

\bibitem[{{Pahl} {et~al.}(2021){Pahl}, {Shapley}, {Steidel}, {Chen}, \&
  {Reddy}}]{pahl21_LyC_clean}
{Pahl}, A.~J., {Shapley}, A., {Steidel}, C.~C., {Chen}, Y., \& {Reddy}, N.~A.
  2021, arXiv e-prints, arXiv:2104.02081

\bibitem[{{Perna} {et~al.}(2015){Perna}, {Brusa}, {Cresci}, {Comastri},
  {Lanzuisi}, {Lusso}, {Marconi}, {Salvato}, {Zamorani}, {Bongiorno},
  {Mainieri}, {Maiolino}, \& {Mignoli}}]{perna15}
{Perna}, M., {Brusa}, M., {Cresci}, G., {et~al.} 2015, \aap, 574, A82

\bibitem[{{Pessa} {et~al.}(2020){Pessa}, {Tejos}, \& {Moya1}}]{PyMUSE2020}
{Pessa}, I., {Tejos}, N., \& {Moya1}, C. 2020, in Astronomical Society of the
  Pacific Conference Series, Vol. 522, Astronomical Data Analysis Software and
  Systems XXVII, ed. P.~{Ballester}, J.~{Ibsen}, M.~{Solar}, \&
  K.~{Shortridge}, 61

\bibitem[{{Pfeffer} {et~al.}(2018){Pfeffer}, {Kruijssen}, {Crain}, \&
  {Bastian}}]{pfeffer18}
{Pfeffer}, J., {Kruijssen}, J.~M.~D., {Crain}, R.~A., \& {Bastian}, N. 2018,
  \mnras, 475, 4309

\bibitem[{{Pignataro} {et~al.}(2021){Pignataro}, {Bergamini}, {Meneghetti},
  {Vanzella}, {Calura}, {Grillo}, {Rosati}, {Angora}, {Brammer}, {Caminha},
  {Mercurio}, {Nonino}, \& {Tozzi}}]{pignataro21}
{Pignataro}, G.~V., {Bergamini}, P., {Meneghetti}, M., {et~al.} 2021, arXiv
  e-prints, arXiv:2106.10286

\bibitem[{{Piqueras} {et~al.}(2019){Piqueras}, {Conseil}, {Shepherd}, {Bacon},
  {Leclercq}, \& {Richard}}]{MPDAF2019}
{Piqueras}, L., {Conseil}, S., {Shepherd}, M., {et~al.} 2019, in Astronomical
  Society of the Pacific Conference Series, Vol. 521, Astronomical Data
  Analysis Software and Systems XXVI, ed. M.~{Molinaro}, K.~{Shortridge}, \&
  F.~{Pasian}, 545

\bibitem[{{Plazas} {et~al.}(2019){Plazas}, {Meneghetti}, {Maturi}, \&
  {Rhodes}}]{Plazas_2019}
{Plazas}, A.~A., {Meneghetti}, M., {Maturi}, M., \& {Rhodes}, J. 2019, \mnras,
  482, 2823

\bibitem[{{Prevot} {et~al.}(1984){Prevot}, {Lequeux}, {Maurice}, {Prevot}, \&
  {Rocca-Volmerange}}]{prevot84}
{Prevot}, M.~L., {Lequeux}, J., {Maurice}, E., {Prevot}, L., \&
  {Rocca-Volmerange}, B. 1984, \aap, 132, 389

\bibitem[{{Reddy} {et~al.}(2016){Reddy}, {Steidel}, {Pettini}, \&
  {Bogosavljevi{\'c}}}]{reddy16}
{Reddy}, N.~A., {Steidel}, C.~C., {Pettini}, M., \& {Bogosavljevi{\'c}}, M.
  2016, \apj, 828, 107

\bibitem[{{Rhoads} {et~al.}(2014){Rhoads}, {Malhotra}, {Richardson},
  {Finkelstein}, {Fynbo}, {McLinden}, \& {Tilvi}}]{rhoads14}
{Rhoads}, J.~E., {Malhotra}, S., {Richardson}, M. L.~A., {et~al.} 2014, \apj,
  780, 20

\bibitem[{{Rigby} {et~al.}(2018{\natexlab{a}}){Rigby}, {Bayliss}, {Chisholm},
  {Bordoloi}, {Sharon}, {Gladders}, {Johnson}, {Paterno-Mahler}, {Wuyts},
  {Dahle}, \& {Acharyya}}]{rigby18_II}
{Rigby}, J.~R., {Bayliss}, M.~B., {Chisholm}, J., {et~al.} 2018{\natexlab{a}},
  \apj, 853, 87

\bibitem[{{Rigby} {et~al.}(2018{\natexlab{b}}){Rigby}, {Bayliss}, {Sharon},
  {Gladders}, {Chisholm}, {Dahle}, {Johnson}, {Paterno-Mahler}, {Wuyts}, \&
  {Kelson}}]{rigby18_I}
{Rigby}, J.~R., {Bayliss}, M.~B., {Sharon}, K., {et~al.} 2018{\natexlab{b}},
  \aj, 155, 104

\bibitem[{{Rigby} {et~al.}(2017){Rigby}, {Johnson}, {Sharon}, {Whitaker},
  {Gladders}, {Florian}, {Lotz}, {Bayliss}, \& {Wuyts}}]{rigby17}
{Rigby}, J.~R., {Johnson}, T.~L., {Sharon}, K., {et~al.} 2017, \apj, 843, 79

\bibitem[{{Rivera-Thorsen} {et~al.}(2019){Rivera-Thorsen}, {Dahle}, {Chisholm},
  {Florian}, {Gronke}, {Rigby}, {Gladders}, {Mahler}, {Sharon}, \&
  {Bayliss}}]{rivera19}
{Rivera-Thorsen}, T.~E., {Dahle}, H., {Chisholm}, J., {et~al.} 2019, Science,
  366, 738

\bibitem[{{Rivera-Thorsen} {et~al.}(2017){Rivera-Thorsen}, {Dahle}, {Gronke},
  {Bayliss}, {Rigby}, {Simcoe}, {Bordoloi}, {Turner}, \& {Furesz}}]{rivera17}
{Rivera-Thorsen}, T.~E., {Dahle}, H., {Gronke}, M., {et~al.} 2017, \aap, 608,
  L4

\bibitem[{{Roberts-Borsani} {et~al.}(2016){Roberts-Borsani}, {Bouwens},
  {Oesch}, {Labbe}, {Smit}, {Illingworth}, {van Dokkum}, {Holden}, {Gonzalez},
  {Stefanon}, {Holwerda}, \& {Wilkins}}]{borsani16}
{Roberts-Borsani}, G.~W., {Bouwens}, R.~J., {Oesch}, P.~A., {et~al.} 2016,
  \apj, 823, 143

\bibitem[{{Romano} {et~al.}(2019){Romano}, {Grazian}, {Giallongo}, {Cristiani},
  {Fontanot}, {Boutsia}, {Fiore}, \& {Menci}}]{romano2019}
{Romano}, M., {Grazian}, A., {Giallongo}, E., {et~al.} 2019, \aap, 632, A45

\bibitem[{{Ryon} {et~al.}(2017){Ryon}, {Gallagher}, {Smith}, {Adamo},
  {Calzetti}, {Bright}, {Cignoni}, {Cook}, {Dale}, {Elmegreen}, {Fumagalli},
  {Gouliermis}, {Grasha}, {Grebel}, {Kim}, {Messa}, {Thilker}, \&
  {Ubeda}}]{ryon17}
{Ryon}, J.~E., {Gallagher}, J.~S., {Smith}, L.~J., {et~al.} 2017, \apj, 841, 92

\bibitem[{{Salpeter}(1955)}]{Salpeter_1955}
{Salpeter}, E.~E. 1955, \apj, 121, 161

\bibitem[{{Schaerer} {et~al.}(2016){Schaerer}, {Izotov}, {Verhamme},
  {Orlitov{\'a}}, {Thuan}, {Worseck}, \& {Guseva}}]{schaerer16}
{Schaerer}, D., {Izotov}, Y.~I., {Verhamme}, A., {et~al.} 2016, \aap, 591, L8

\bibitem[{{Sharon} {et~al.}(2020){Sharon}, {Bayliss}, {Dahle}, {Dunham},
  {Florian}, {Gladders}, {Johnson}, {Mahler}, {Paterno-Mahler}, {Rigby},
  {Whitaker}, {Akhshik}, {Koester}, {Murray}, {Remolina Gonz{\'a}lez}, \&
  {Wuyts}}]{sharon20}
{Sharon}, K., {Bayliss}, M.~B., {Dahle}, H., {et~al.} 2020, \apjs, 247, 12

\bibitem[{{Smit} {et~al.}(2015){Smit}, {Bouwens}, {Franx}, {Oesch}, {Ashby},
  {Willner}, {Labb{\'e}}, {Holwerda}, {Fazio}, \& {Huang}}]{smit15neb}
{Smit}, R., {Bouwens}, R.~J., {Franx}, M., {et~al.} 2015, \apj, 801, 122

\bibitem[{{Smit} {et~al.}(2016){Smit}, {Bouwens}, {Labb{\'e}}, {Franx},
  {Wilkins}, \& {Oesch}}]{smit16neb}
{Smit}, R., {Bouwens}, R.~J., {Labb{\'e}}, I., {et~al.} 2016, \apj, 833, 254

\bibitem[{{Smit} {et~al.}(2014){Smit}, {Bouwens}, {Labb{\'e}}, {Zheng},
  {Bradley}, {Donahue}, {Lemze}, {Moustakas}, {Umetsu}, {Zitrin}, {Coe},
  {Postman}, {Gonzalez}, {Bartelmann}, {Ben{\'\i}tez}, {Broadhurst}, {Ford},
  {Grillo}, {Infante}, {Jimenez-Teja}, {Jouvel}, {Kelson}, {Lahav}, {Maoz},
  {Medezinski}, {Melchior}, {Meneghetti}, {Merten}, {Molino}, {Moustakas},
  {Nonino}, {Rosati}, \& {Seitz}}]{smit14neb}
{Smit}, R., {Bouwens}, R.~J., {Labb{\'e}}, I., {et~al.} 2014, \apj, 784, 58

\bibitem[{{Steidel} {et~al.}(2018){Steidel}, {Bogosavljevi{\'c}}, {Shapley},
  {Reddy}, {Rudie}, {Pettini}, {Trainor}, \& {Strom}}]{steidel18}
{Steidel}, C.~C., {Bogosavljevi{\'c}}, M., {Shapley}, A.~E., {et~al.} 2018,
  \apj, 869, 123

\bibitem[{{Steidel} {et~al.}(2001){Steidel}, {Pettini}, \&
  {Adelberger}}]{steidel2001}
{Steidel}, C.~C., {Pettini}, M., \& {Adelberger}, K.~L. 2001, \apj, 546, 665

\bibitem[{{Stuik} {et~al.}(2006){Stuik}, {Bacon}, {Conzelmann}, {Delabre},
  {Fedrigo}, {Hubin}, {Le Louarn}, \& {Str{\"o}bele}}]{GALACSI06}
{Stuik}, R., {Bacon}, R., {Conzelmann}, R., {et~al.} 2006, \nar, 49, 618

\bibitem[{{Tang} {et~al.}(2021){Tang}, {Stark}, {Chevallard}, {Charlot},
  {Endsley}, \& {Congiu}}]{mengtao21}
{Tang}, M., {Stark}, D.~P., {Chevallard}, J., {et~al.} 2021, \mnras, 503, 4105

\bibitem[{{Trebitsch} {et~al.}(2017){Trebitsch}, {Blaizot}, {Rosdahl},
  {Devriendt}, \& {Slyz}}]{trebitsch2017}
{Trebitsch}, M., {Blaizot}, J., {Rosdahl}, J., {Devriendt}, J., \& {Slyz}, A.
  2017, \mnras, 470, 224

\bibitem[{{van der Walt} {et~al.}(2011){van der Walt}, {Colbert}, \&
  {Varoquaux}}]{NUMPY2011}
{van der Walt}, S., {Colbert}, S.~C., \& {Varoquaux}, G. 2011, Computing in
  Science and Engineering, 13, 22

\bibitem[{{Vanzella} {et~al.}(2019){Vanzella}, {Calura}, {Meneghetti},
  {Castellano}, {Caminha}, {Mercurio}, {Cupani}, {Rosati}, {Grillo}, {Gilli},
  {Mignoli}, {Fiorentino}, {Arcidiacono}, {Lombini}, \& {Cortecchia}}]{vanz19}
{Vanzella}, E., {Calura}, F., {Meneghetti}, M., {et~al.} 2019, \mnras, 483,
  3618

\bibitem[{{Vanzella} {et~al.}(2017){Vanzella}, {Calura}, {Meneghetti},
  {Mercurio}, {Castellano}, {Caminha}, {Balestra}, {Rosati}, {Tozzi}, {De
  Barros}, {Grazian}, {D'Ercole}, {Ciotti}, {Caputi}, {Grillo}, {Merlin},
  {Pentericci}, {Fontana}, {Cristiani}, \& {Coe}}]{vanz_paving}
{Vanzella}, E., {Calura}, F., {Meneghetti}, M., {et~al.} 2017, \mnras, 467,
  4304

\bibitem[{{Vanzella} {et~al.}(2020{\natexlab{a}}){Vanzella}, {Caminha},
  {Calura}, {Cupani}, {Meneghetti}, {Castellano}, {Rosati}, {Mercurio}, {Sani},
  {Grillo}, {Gilli}, {Mignoli}, {Comastri}, {Nonino}, {Cristiani},
  {Giavalisco}, \& {Caputi}}]{vanz_sunburst}
{Vanzella}, E., {Caminha}, G.~B., {Calura}, F., {et~al.} 2020{\natexlab{a}},
  \mnras, 491, 1093

\bibitem[{{Vanzella} {et~al.}(2021){Vanzella}, {Caminha}, {Rosati}, {Mercurio},
  {Castellano}, {Meneghetti}, {Grillo}, {Sani}, {Bergamini}, {Calura},
  {Caputi}, {Cristiani}, {Cupani}, {Fontana}, {Gilli}, {Grazian}, {Gronke},
  {Mignoli}, {Nonino}, {Pentericci}, {Tozzi}, {Treu}, {Balestra}, \&
  {Dijkstra}}]{vanz_mdlf}
{Vanzella}, E., {Caminha}, G.~B., {Rosati}, P., {et~al.} 2021, \aap, 646, A57

\bibitem[{{Vanzella} {et~al.}(2016){Vanzella}, {de Barros}, {Vasei}, {Alavi},
  {Giavalisco}, {Siana}, {Grazian}, {Hasinger}, {Suh}, {Cappelluti}, {Vito},
  {Amorin}, {Balestra}, {Brusa}, {Calura}, {Castellano}, {Comastri}, {Fontana},
  {Gilli}, {Mignoli}, {Pentericci}, {Vignali}, \& {Zamorani}}]{vanz_ion2}
{Vanzella}, E., {de Barros}, S., {Vasei}, K., {et~al.} 2016, \apj, 825, 41

\bibitem[{{Vanzella} {et~al.}(2020{\natexlab{b}}){Vanzella}, {Meneghetti},
  {Caminha}, {Castellano}, {Calura}, {Rosati}, {Grillo}, {Dijkstra}, {Gronke},
  {Sani}, {Mercurio}, {Tozzi}, {Nonino}, {Cristiani}, {Mignoli}, {Pentericci},
  {Gilli}, {Treu}, {Caputi}, {Cupani}, {Fontana}, {Grazian}, \&
  {Balestra}}]{vanz_popiii}
{Vanzella}, E., {Meneghetti}, M., {Caminha}, G.~B., {et~al.}
  2020{\natexlab{b}}, \mnras, 494, L81

\bibitem[{{Vanzella} {et~al.}(2020{\natexlab{c}}){Vanzella}, {Meneghetti},
  {Pastorello}, {Calura}, {Sani}, {Cupani}, {Caminha}, {Castellano}, {Rosati},
  {D'Odorico}, {Cristiani}, {Grillo}, {Mercurio}, {Nonino}, {Brammer}, \&
  {Hartman}}]{vanz20_laser}
{Vanzella}, E., {Meneghetti}, M., {Pastorello}, A., {et~al.}
  2020{\natexlab{c}}, \mnras, 499, L67

\bibitem[{{Vanzella} {et~al.}(2018){Vanzella}, {Nonino}, {Cupani},
  {Castellano}, {Sani}, {Mignoli}, {Calura}, {Meneghetti}, {Gilli}, {Comastri},
  {Mercurio}, {Caminha}, {Caputi}, {Rosati}, {Grillo}, {Cristiani}, {Balestra},
  {Fontana}, \& {Giavalisco}}]{vanz18}
{Vanzella}, E., {Nonino}, M., {Cupani}, G., {et~al.} 2018, \mnras, 476, L15

\bibitem[{{Vargas-Salazar} {et~al.}(2020){Vargas-Salazar}, {Oey}, {Barnes},
  {Chen}, {Castro}, {Kratter}, \& {Faerber}}]{Vargas_Salazar20}
{Vargas-Salazar}, I., {Oey}, M.~S., {Barnes}, J.~R., {et~al.} 2020, \apj, 903,
  42

\bibitem[{{Wang} {et~al.}(2021){Wang}, {Heckman}, {Amor{\'\i}n}, {Borthakur},
  {Chisholm}, {Ferguson}, {Flury}, {Giavalisco}, {Grazian}, {Hayes}, {Henry},
  {Jaskot}, {Ji}, {Makan}, {McCandliss}, {Oey}, {{\"O}stlin}, {Saldana-Lopez},
  {Schaerer}, {Thuan}, {Worseck}, \& {Xu}}]{Wang_[SII]_21}
{Wang}, B., {Heckman}, T.~M., {Amor{\'\i}n}, R., {et~al.} 2021, arXiv e-prints,
  arXiv:2104.03432

\bibitem[{{Williams} {et~al.}(2021){Williams}, {Spilker}, {Whitaker},
  {Dav{\'e}}, {Woodrum}, {Brammer}, {Bezanson}, {Narayanan}, \&
  {Weiner}}]{williams21}
{Williams}, C.~C., {Spilker}, J.~S., {Whitaker}, K.~E., {et~al.} 2021, \apj,
  908, 54

\bibitem[{{Wise} {et~al.}(2014){Wise}, {Demchenko}, {Halicek}, {Norman},
  {Turk}, {Abel}, \& {Smith}}]{wise2014}
{Wise}, J.~H., {Demchenko}, V.~G., {Halicek}, M.~T., {et~al.} 2014, \mnras,
  442, 2560

\bibitem[{{Wofford} {et~al.}(2014){Wofford}, {Leitherer}, {Chandar}, \&
  {Bouret}}]{wofford14}
{Wofford}, A., {Leitherer}, C., {Chandar}, R., \& {Bouret}, J.-C. 2014, \apj,
  781, 122

\bibitem[{{Worseck} {et~al.}(2014){Worseck}, {Prochaska}, {O'Meara}, {Becker},
  {Ellison}, {Lopez}, {Meiksin}, {M{\'e}nard}, {Murphy}, \&
  {Fumagalli}}]{worseck14}
{Worseck}, G., {Prochaska}, J.~X., {O'Meara}, J.~M., {et~al.} 2014, \mnras,
  445, 1745

\bibitem[{{Wuyts} {et~al.}(2013){Wuyts}, {F{\"o}rster Schreiber}, {Nelson},
  {van Dokkum}, {Brammer}, {Chang}, {Faber}, {Ferguson}, {Franx}, {Fumagalli},
  {Genzel}, {Grogin}, {Kocevski}, {Koekemoer}, {Lundgren}, {Lutz}, {McGrath},
  {Momcheva}, {Rosario}, {Skelton}, {Tacconi}, {van der Wel}, \&
  {Whitaker}}]{wuyts13}
{Wuyts}, S., {F{\"o}rster Schreiber}, N.~M., {Nelson}, E.~J., {et~al.} 2013,
  \apj, 779, 135

\bibitem[{{Zanella} {et~al.}(2015){Zanella}, {Daddi}, {Le Floc'h}, {Bournaud},
  {Gobat}, {Valentino}, {Strazzullo}, {Cibinel}, {Onodera}, {Perret}, {Renaud},
  \& {Vignali}}]{zanella15}
{Zanella}, A., {Daddi}, E., {Le Floc'h}, E., {et~al.} 2015, \nat, 521, 54

\bibitem[{{Zanella} {et~al.}(2019){Zanella}, {Le Floc'h}, {Harrison}, {Daddi},
  {Bernhard}, {Gobat}, {Strazzullo}, {Valentino}, {Cibinel}, {S{\'a}nchez
  Almeida}, {Kohandel}, {Fensch}, {Behrendt}, {Burkert}, {Onodera}, {Bournaud},
  \& {Scholtz}}]{zanella19}
{Zanella}, A., {Le Floc'h}, E., {Harrison}, C.~M., {et~al.} 2019, \mnras, 489,
  2792

\end{thebibliography}

\begin{appendix}

\section{\ha\ and \oiiiv\ emissions from star-forming knots}
\label{ha}

\begin{figure*}
        \centering
        \includegraphics[width=\linewidth]{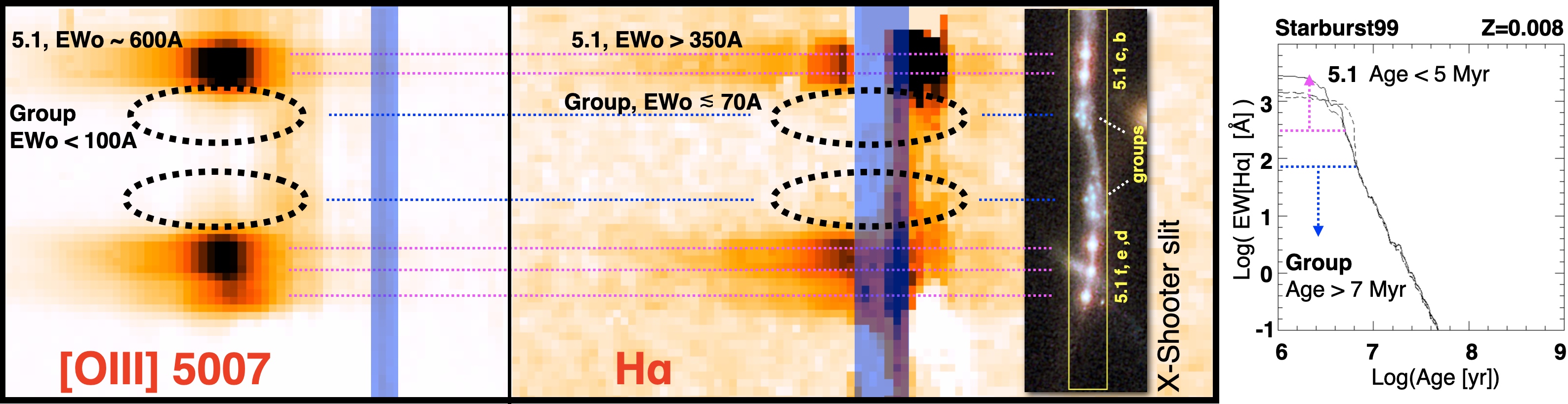}
        \caption{Main panel (thick line box): the \oiiiv\ (left) and the \ha\ (right) two-dimensional zoomed spectra extracted from X-Shooter observations at spectral resolution R~$\simeq5600$. The slit orientation is indicated with the yellow box in the inset showing the HST color image. The magenta dotted lines mark the position of the knots 5.1b, c and 5.1d, e, f over the spatial direction along the slit. The blue dotted lines  show the expected positions of the region collecting the group of fainter knots, named "group" in the figure (see also Figure~\ref{pano}). \ha\ and \oiiiv\ clearly emerge from knots 5.1b,c,d,e,f, conversely the same lines are not detected for the two mirrored "groups" at the given depth. The inferred rest-frame equivalent widths are also quoted. The transparent vertical blue stripes mark the position of night sky emission lines. The right panel shows the
        \ha\ equivalent width as a function of time from a {\rm Starburst99} model of an instantaneous burst and Z=0.008Z$_{\odot}$. Magenta and blue dotted arrows indicate the case of the knot 5.1 and "Group," respectively. Knot 5.1 is younger than 5 Myr (in agreement with the 3Myr old age derived by \citealt{chisholm19}), while the complex "group" is likely older than 7 Myr.}
        \label{zoomknots}
\end{figure*}

X-Shooter observations have been performed during April-August 2019 (Prog. 0103.A-0688, PI Vanzella). The two arcs I and II have been targeted with two slits oriented along the tangential stretch.
The data reduction is described in \citet{vanz20_laser}.
Briefly, no dithering has been used since the arcs develop along the spatial direction probed with the slits. Suitable free windows without bright objects have been included such that sky was modeled with polynomial fitting and subtracted. 
Here we focus on a specific region of arc I where two mirrored regions containing 10 SF knots each have been observed (see Figure~\ref{zoomknots}). In particular knot 5.1c,b and 5.1d,e,f, as well as the groups "c" and "d" of blue knots also shown in Figure~\ref{pano} (5.4, 5.5, 5.6, 5.7, 5.3, 5.9, 5.2, and 5.11). Figure~\ref{zoomknots} shows the spectrum at \oiiiv\ and \ha\ wavelengths. Both lines emerge prominently from the YMC 5.1, while for the groups we can set only upper limits on their line fluxes and equivalent widths. As a test case, we verified the consistency between the \oiiiv\ line flux inferred from VLT/X-Shooter and the photometric jump observed in the SED fitting performed on the same aperture probing 5.1b,c (as indicated in Figure~\ref{sed}). The 0.6 magnitude boost in the H band is fully consistent with the inferred \oiiiv\ equivalent width, also including the \oiiiiv\ and \hb. What is inferred from the spectrum is slightly fainter than what observed with Hubble photometry because of slit losses.

Similarly, the \ha\ emission is also prominent from knot 5.1. It is worth noting that the multiple images within the slit allow us to formally quintuple the integration time on 5.1, and duplicate it over the groups "c" and "d". The resulting equivalent width of \ha\ is higher than 350\AA\ rest-frame for knot 5.1 and lower than 100\AA\ for the group of knots. Such values under the assumption of an instantaneous burst of star formation imply ages younger than 5 Myr for the YMC knot 5.1 and older than $\simeq$~7 Myr on average for the group of knots (see rightmost side of Figure~\ref{ha}). Such a lower limit on the age for the group of knots is the value assumed in the text and reported in Table~\ref{knots}. The case of YMC 5.1 is fully consistent with the age inferred from \citet{chisholm19}
of 2.9 Myr (from Starburst99 modeling).

\section{Dynamical mass of the LyC star cluster}
\label{dyn_mass}

\begin{figure}
        \centering
        \includegraphics[width=\linewidth]{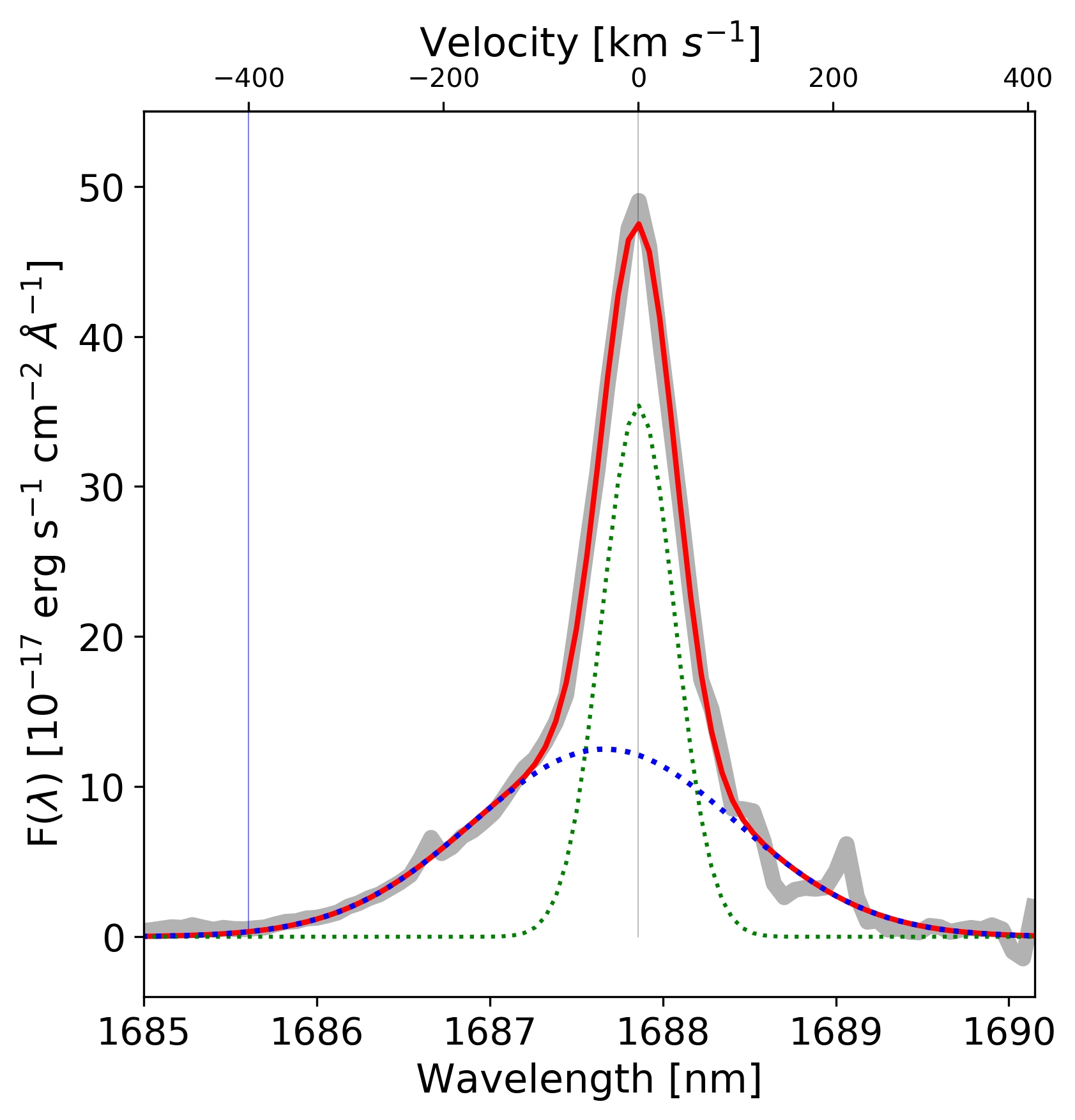}
        \caption{\oiiiv\ line emerging from the LyC knot 5.1 (gray band) and the best-fit solution made of the sum of two Gaussian components (red line), a broad one tracing the outflowing ionized gas (blue dotted) and the other tracing the core of the line; the latter arises from the \hii\ region surrounding the star cluster and tracing the redshift of the stellar component (green dotted). On top axis, the velocity is referred to the peak position of the line. The clear blue tail extends up to $-$400 \kms.}
        \label{OIIIline}
\end{figure}

The \oiiiv\ line is detected at S/N$>$100, it is free from significant sky emission and shows an evident asymmetric profile with a blue tail (see Figure~\ref{OIIIline}). A  fit with a double Gaussian component has been performed and reproduces well the observed profile, suggesting there is outflowing ionized gas along the line of sight (broad component) and the emission associated to the \hii\ region close to the stellar component and tracing the systemic redshift (core of the line). Such an outflow is consistent with what was already observed by \citet{rivera17}
based on absorption lines of Silicon IV. Here we focus on the core of the line. The resulting velocity dispersion of such an emission from the fit is $\sigma_v = 37 \pm 5$ \kms\ (the resolution element in the NIR arm is $d\lambda \simeq 53$ \kms, or $\sigma_v \simeq 22$ \kms).
Adopting the inferred $\sigma_v$ and the known effective radius $R_{\tt eff} \simeq 8$ pc of the YMC (5.1), the dynamical mass can be calculated following \citet{rhoads14} as Mdyn = $4\sigma^2 R_{\tt eff} / G$ (see also \citealt{vanz_paving}), which turns out to be $(1.0 \pm 0.2) \times 10^{7}$ \msun. Therefore, both the photometric and dynamical masses are compatible, with a ratio close to 1. As discussed in the text, the presence of the magnification factors in the numerator and denominator implies the ratio between stellar and virial masses weakly depends on lensing amplification.

\section{MUSE Narrow Field Mode spectroscopic observations}
\label{appendix_NFM}

It is worth reporting here on the exceptional MUSE observation in the Narrow Field Mode (PI Vanzella) on arc IV, which provides spectroscopy and imaging at resolution smaller than 80 mas and better than HST blue bands. 
The proximity of a sufficiently bright star (H~$<14$ Vega,  also dubbed as natural guide star, NGS) within $3.35''$ angular separation from the targeted counterarc (arc IV) allowed for dedicated nearly diffraction-limited observations with MUSE \citep[][]{Bacon_MUSE} Narrow Field Mode, offered by the GALACSI module implementing the laser tomography AO correction \citep[][]{GALACSI06}. In this configuration, MUSE provides a field of view of $7.5'' \times 7.5''$ sampled at $0.025''$/pixel scale. The relatively high (best) airmass of $\sim 1.7$ for the target is typically not ideal for applying extreme AO corrections, however, we requested proper weather conditions as suggested in the manual for such airmass (seeing $< 0.6''$ at zenith). A total of 2.6h divided into two observing blocks (OB) of 3532s each on science (each OB split into $883 \times 4$) have been performed during April 29 2019 with the exceptional average seeing conditions of $0.4''$ and producing a Strehl on the NGS of $\simeq 20$\%. Data reduction have been performed following the prescription described in \citep[][]{Caminha_macs0416,vanz_mdlf}.

\begin{figure*}
        \centering
        \includegraphics[width=\linewidth]{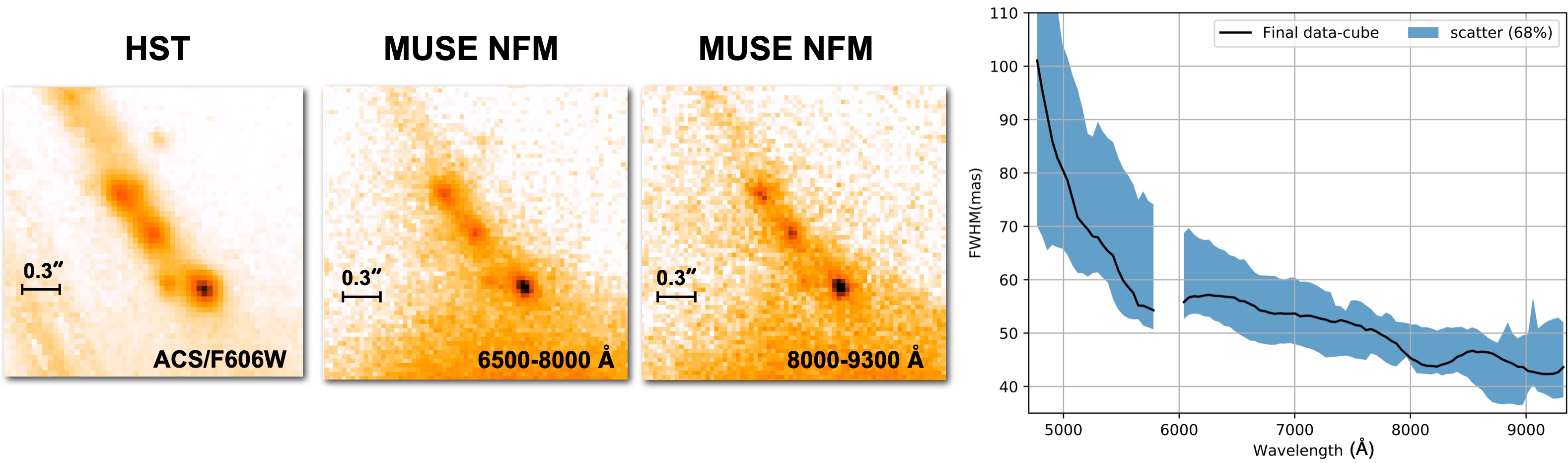}
        \caption{Counter arc IV observed with HST ACS/F606W (left) and VLT/MUSE NFM after collapsing the datacube along the indicated wavelength intervals (middle and right images). The ground-based AO-assisted observations of arc IV outperform those from space (see narrow-band imaging after collapsing the cube in the wavelength ranges 6500-8000\AA\  and 8000-9000\AA), providing further evidences for the compactness of the star-forming region and in particular the LyC star cluster 5.1 (in this case, the image 5.1n has been targeted). 
        The right most figure shows the FWHM of the OGS (On-axis Guide Source), measured within a 1.5\arcsec\ region, as a function of the wavelength. The black line show the measurements in the final stacked data-cube, whereas the blue region indicates the scatter (68\% confidence level) of the measurements in each single exposure. The FWHM is $< 60$ mas at $\lambda > 7000$\AA\ and we assume a similar resolution is achieved on arc IV (middle).}
        \label{NFM}
\end{figure*}

Figure~\ref{NFM} shows the FWHM of the Moffat profile fitted on the NGS (on-axis) star as a function of the wavelength. Remarkably, the FWHM decreases at values lower than 60 mas at $\lambda > 6000$\AA, approaching $50-55$ mas over the range $7000-9250$\AA. The same figure also shows the comparison on arc IV, between HST/ACS F606W and MUSE collapsed image in the range $6500-8000$\AA\ and $8000-9300$\AA. Even though the off-axis ($\sim 3''$ apart) FWHM is probably worse than the on-axis one on the NGS, MUSE with its FWHM~$ < 100$ mas appears to outperform HST F606W ($\sim 100$ mas) resolutions and allow us to probe smaller physical scales, at least in this specific arc.

Despite arc IV is not the most magnified, an upper limit on the size of the star-forming knot 5.1 of 40 pc (adopting FWHM~$< 60$ mas) represents a remarkable technical achievement, considering the relatively high airmass (1.7). If, on the one hand, this MUSE-NFM shows its exceptional technical capabilities, which matches the HST imaging; on the other hand the poor flexibility (in terms of sky coverage) due to the need for a very close (within $3.35''$) and bright (H~$<14$ Vega magnitude) natural guide star makes the observations of the most magnified arcs I or II inaccessible. Extreme AO-correction with much wider sky coverage is therefore essential for the future VLT instrumentation, such as MAVIS (7 mas/pixel, FWHM~$\sim 25$ mas in the optical bands), which will provide spatial details down to $1-3$ pc/pix on arcs I or II (see Sect.~\ref{future}).

\end{appendix}

\end{document}